\documentclass[prl,reprint,aps,amsmath,amssymb,twocolumn,superscriptaddress,longbibliography]{revtex4-1}
\usepackage{graphicx,bm,color,appendix}
\usepackage[colorlinks,bookmarks=true,citecolor=blue,linkcolor=blue,urlcolor=blue]{hyperref}
\usepackage{times}
\usepackage{amsmath,amssymb}
\usepackage{comment}
\usepackage{appendix}

\graphicspath{{./fig/}}

\usepackage{xcolor}
\definecolor{orange}{rgb}{1,0.5,0}

\newcommand{\tj}[6]{ \begin{pmatrix}
   #1 & #2 & #3 \\
   #4 & #5 & #6 
  \end{pmatrix}}

\usepackage[normalem]{ulem}
\usepackage{comment}

\begin{document}

\title{Solving Conformal Defects in 3D Conformal Field Theory using Fuzzy Sphere Regularization}

\author{Liangdong Hu}
\affiliation{Westlake Institute of Advanced Study,	 Westlake University, Hangzhou 310024, China }

\author{Yin-Chen He}
\email{yhe@perimeterinstitute.ca}
\affiliation{Perimeter Institute for Theoretical Physics, Waterloo, Ontario N2L 2Y5, Canada}

\author{W. Zhu}
\email{zhuwei@westlake.edu.cn}
\affiliation{Westlake Institute of Advanced Study, Westlake University, Hangzhou 310024, China }

\begin{abstract}
Defects in conformal field theory (CFT) are of significant theoretical and experimental importance. The presence of defects theoretically enriches the structure of the CFT, but at the same time, it makes it more challenging to study, especially in dimensions higher than two.
Here, we demonstrate that the recently-developed theoretical scheme, \textit{fuzzy (non-commutative) sphere regularization}, provides a powerful lens through which one can dissect the defect of 3D CFTs in a transparent way. As a notable example, we study the magnetic line defect of 3D Ising CFT and clearly demonstrate that it flows to a conformal defect fixed point. We have identified 6 low-lying defect primary operators, including the displacement operator, and accurately extract their scaling dimensions through the state-operator correspondence.
Moreover, we also compute one-point bulk correlators and two-point bulk-defect correlators, which show great agreement with predictions of defect conformal symmetry, and from which we extract various bulk-defect operator product expansion coefficients. 
Our work demonstrates that the fuzzy sphere offers a powerful tool for exploring the rich physics in 3D defect CFTs.
\end{abstract}

\date{\today}

\maketitle

Defects, as well as their special case--boundaries, are fundamental elements that inevitably exist in nearly all realistic physical systems. Historically, research on defects has played a pivotal role in shaping modern theoretical physics. This includes contributions to the theory of the renormalization group (RG)~\cite{Wilson1975}, studies of topological phases~\cite{anyon_RMP_2008,Hasan2010Review,Qi2011Reivew}, investigations into the confinement of gauge theories~\cite{Wilson1974Confinement,hooft1978phase}, explorations of quantum gravity~\cite{Maldacena_AdSCFT}, and advancements in the understanding of quantum entanglement~\cite{HOLZHEY1994,calabrese2004entanglement}. 
An important instance to study defects is in the context of conformal field theory (CFT)~\cite{yellowbook,Cardy_book}, where one considers the situation of deforming a CFT with interactions living on a sub-dimensional defect. The defect may trigger an RG flow towards a non-trivial infrared (IR) fixed point, which can still have an emergent conformal symmetry defined on the space-time dimensions of the defect~\cite{CARDY1984surface,cardy1989boundary,cardy1991bulkboundary,Diehl1981Field-theoretical,McAvity1993EMT,MCAVITY1995}. The theory describing such a conformal defect is called a defect CFT (dCFT) (see Refs.~\cite{Billo2013,Billo2016defect} for recent discussions). Understanding dCFTs is an important step in comprehending CFTs in nature, as most experimental realizations of CFTs necessarily accompany defects (and boundaries). Moreover, dCFTs have a non-trivial interplay with the bulk CFTs, and knowledge of the former will advance the understanding of the latter. For example, the two-point correlators of bulk operators in dCFT constrain and encode the conformal data of the bulk CFT~\cite{Liendo2013}, similar to the well-known story of four-point correlators of a bulk CFT.  

dCFTs are typically richer and more intricate than their bulk CFT counterparts. On one hand, for a given bulk CFT, there exist multiple (even potentially infinite) distinct dCFTs, and their classification remains an open challenge. 
On the other hand, breaking of the full conformal symmetry group into a subgroup renders the study of dCFTs more challenging, as the space-time conformal symmetry becomes less restrictive, making modern approaches like the conformal bootstrap program~\cite{RMP_CB} less powerful~\cite{Liendo2013,Gaiotto2014,Gliozzi2015Boundary,Padayasi2022extraordinary,Gimenez-Grau2022Bootstrapping}. Notably, most of the well-established results concerning dCFTs are confined to 2D CFTs, including the seminal results on the boundary operator contents~\cite{cardy1989boundary} and RG flow~\cite{Affleck1991Universal}, thanks to the special integrability property of 2D CFTs. In comparison, higher-dimensional CFTs pose greater difficulties, and the knowledge of dCFTs in dimensions beyond two is rather limited. Current studies of dCFTs mainly revolve around perturbative RG computations~\cite{Vojta2000Impurity,Metlitski2020Boundary,Liu2021Magnetic,Krishnan2023plane,Aharony2023Phases,Hanke2000,Andrea2014,Cuomo2022magneticDefect} and Monte Carlo simulations of lattice models \cite{Billo2013,Allais2014,Assaad2017,Toldin2022Boundary}. An important progress made recently is the non-perturbative proof of RG monotonic g-theorem in 3D and higher dimensions~\cite{Cuomo2022gtheorem,Casini2023Entropic}, generalizing the original result in 2D~\cite{Affleck1991Universal,Friedan2004Boundary,Casini20162Dg}.  

In the context of dCFTs, many important questions remain to be answered, ranging from basic inquiries such as the existence of conformal defect fixed points to more advanced queries concerning the infrared properties of dCFTs, including their conformal data such as critical exponents.
The central aim of this paper is to develop an efficient tool for the non-perturbative analysis of 3D dCFTs. Specifically, we extend the success of the recently proposed fuzzy sphere regularization~\cite{ZHHHH2022} from bulk CFTs~\cite{ZHHHH2022,hu2023operator,Han2023Conformal,Zhou2023SO5} to the realm of dCFTs.
As a concrete example, we explore the properties of the 3D Ising CFT in the presence of a magnetic line defect~\cite{Hanke2000,Andrea2014,Cuomo2022magneticDefect,Nishioka2023,Allais2014,Assaad2017,Bianchi2023Analytic,Gimenez-Grau2022Probing,Pannell2023Line}. We directly demonstrate that this line defect indeed flows to an attractive conformal fixed point, and we identify 6 low-lying defect primary operators with their scaling dimensions extracted through the state-operator correspondence. Furthermore, we study the one-point bulk primary correlators and the two-point bulk-defect correlators, both of which are fixed by conformal invariance, up to a set of operator product expansion (OPE) coefficients. As far as we know, most of conformal data of dCFT reported here have never been studied before.
In this context, our paper not only presents a comprehensive set of results concerning the magnetic line defect in the 3D Ising CFT, but also lays the foundation for further exploration of 3D dCFTs using the fuzzy sphere regularization technique.

\section*{Results}
\subsection*{Conformal defect and radial quantization}
\begin{figure}[b]
\includegraphics[width=0.95\linewidth]{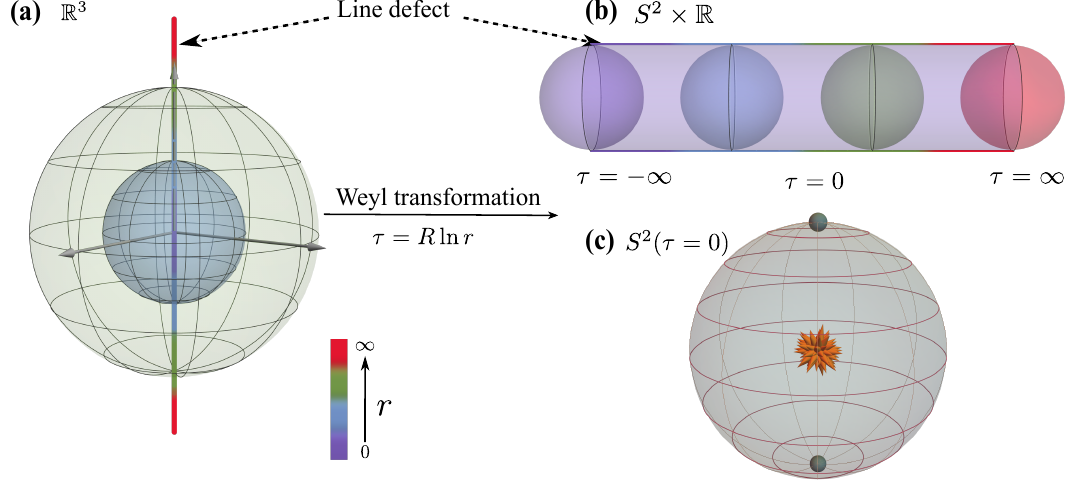}
\caption{\textbf{Schematic plot of the defect in 3D}. 
Through a Weyl transformation, (a) Euclidean flat space-time $\mathbb R^3$  is mapped to (b) the cylinder manifold $S^{2}\times \mathbb R$. The line defect before and after the Weyl transformation are shown by the colored line.
(c) The $0+1$-D impurities (cyan point) located at the north and south pole on two-dimensional sphere $S^{2}$ in the radial quantization, where the flux at the center represents the magnetic monopole defined in the fuzzy sphere model.   
}
\label{fig:drawing_defect}
\end{figure}

We consider a 3D CFT deformed by a $p$-dimensional defect, described by the Hamiltonian 
\begin{align} \label{eq:Ham_dCFT}
	H_{CFT} + h \int d^p r \mathcal{O}(r).
\end{align} 
Examples include the line defect ($p=1$, see Fig.~\ref{fig:drawing_defect}(a)) and the plane defect ($p=2$). If the defect is not screened in the IR, the system will flow into a non-trivial fixed point that breaks the original conformal symmetry $SO(4,1)$ of $H_{CFT}$. Furthermore, if the non-trivial fixed point is still conformal, such a defect is called a conformal defect described by a dCFT. For such a dCFT, the original conformal group is broken down to a smaller subgroup $SO(p+1,1) \times SO(3-p)$~\cite{MCAVITY1995,Billo2016defect,Billo2013}, where $SO(p+1,1)$ is the conformal symmetry of the defect, and $SO(3-p)$ is the rotation symmetry around the defect that acts as a global symmetry on the defect.

\begin{figure}[b]
\includegraphics[width=0.97\linewidth]{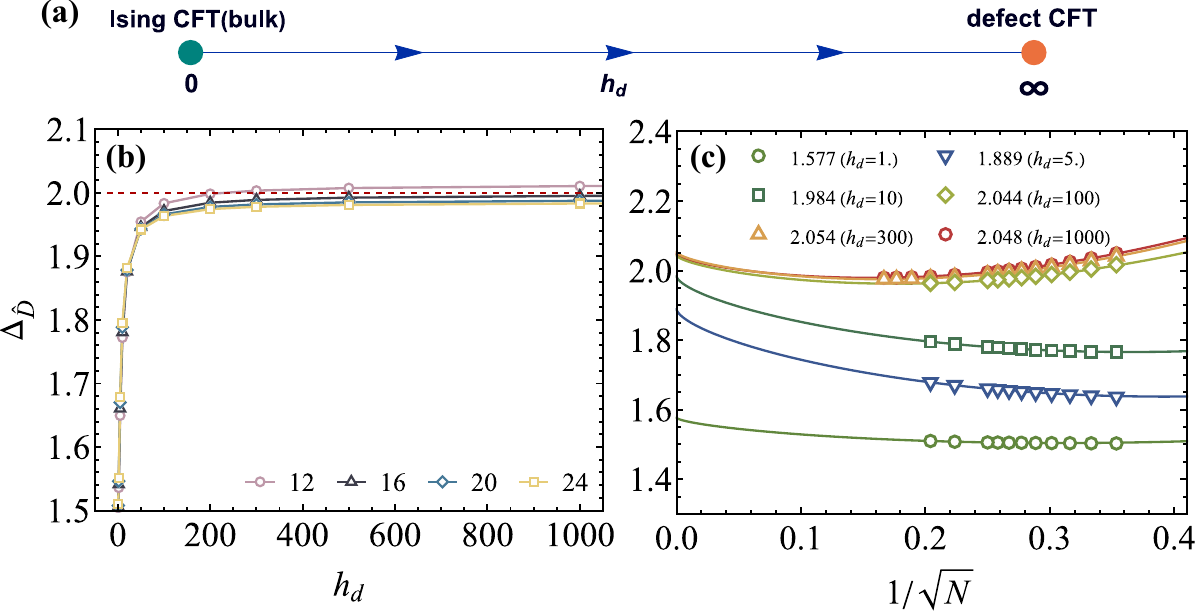}
\caption{\textbf{Defect induced attractive fixed point}. (a) Schematic plot of the RG diagram. (b) The scaling dimension of displacement operator $\Delta_{\hat D}$ as a function of defect strength $h_d$. Different colored symbols represent the results based on various system sizes. (c) Finite-size extrapolation of $\Delta_{\hat D}$ for various $h_d$ \cite{SM}. A sufficient large $h_d$ gives almost identical $\Delta_{\hat D}\approx 2$, supporting an attractive RG fixed point at $h_d=\infty$. Different colored symbols represent the results based on various defect strength $h_d$.
}
\label{fig:displacement_flow}
\end{figure}

A dCFT possesses a richer structure compared to its bulk counterpart. Firstly, there is a set of operators living on the defect, forming representations of the defect conformal group $SO(p+1,1)$. Furthermore, there are non-trivial correlators between bulk operators and defect operators. (Hereafter, we follow the usual convention and denote the defect operator with a hat $\hat O$, while the bulk operator is represented as $O$ without a hat.)
The simplest example is that the bulk primary operator gets a non-vanishing one-point correlator, which is in sharp contrast to the bulk CFT \cite{Billo2013,Billo2016defect,MCAVITY1995}:
\begin{equation}\label{eq:1pt}
\langle O_1(x) \rangle = \frac{a_{O_1}}{|x_{\perp}|^{\Delta_{1}}}.
\end{equation}
Here, $|x_\perp|$ is the perpendicular distance from the bulk operator to the defect, $\Delta_1$ is the scaling dimension of $O_1$, and
$a_{O_1}$ is an operator-dependent universal number (we consider the case of $O_1$ to be a Lorentz scalar).
Moreover, we can consider a bulk-defect two-point (scalar-scalar) correlator defined as \cite{Billo2013,Billo2016defect,MCAVITY1995}:
\begin{equation}\label{eq:2pt}
\langle O_1(x) \hat O_2(0) \rangle = \frac{b_{O_1 \hat O_2}}{|x_\perp|^{\Delta_1- \hat \Delta_2} |x|^{2\hat \Delta_2}},
\end{equation}
where $b_{O_1 \hat O_2}$ is the bulk-defect OPE coefficient. Interestingly, the bulk two-point correlator already becomes non-trivial, and its functional form cannot be completely fixed by the conformal symmetry.

Similar to the bulk CFT, we consider the radial quantization of a dCFT. Specifically, we first foliate the Euclidean space $\mathbb{R}^3$ using spheres $S^2$ with their origins situated on the defect, as illustrated in Fig.~\ref{fig:drawing_defect}(a). Next, we can perform a Weyl transformation to map $\mathbb{R}^3$ to a cylinder $S^2\times \mathbb{R}$, and the $p$-dimensional defect transforms into a defect intersecting the cylinder. For instance, as shown in Fig. \ref{fig:drawing_defect}, the Weyl transformation maps a line defect ($p=1$) in $\mathbb{R}^3$ to $0+1$D point impurities located at the north and south poles of the sphere $S^2$, forming two continuous line cuts along the time direction from $\tau=-\infty$ to $\tau=\infty$. Similarly, a plane defect ($p=2$) in $\mathbb{R}^3$ will be mapped to a $1+1$D defect with its spatial component located on the equator of the sphere $S^2$.

Akin to the state-operator correspondence in bulk CFT~\cite{Cardy1984,Cardy1985}, we have a one-to-one correspondence between the defect operators and the eigenstates of the dCFT quantum Hamiltonian on $S^2 \times \mathbb{R}$, where energy gaps of these states are proportional to the scaling dimensions $\hat{\Delta}_n$ of the defect operators:
\begin{equation} \label{eq:correspondence}
E_n - E_0 = \frac{v}{R} \hat{\Delta}_n.
\end{equation}
Here, $E_0$ denotes the ground state energy of the defect Hamiltonian, $R$ represents the sphere radius, and $v$ is a model-dependent non-universal velocity that corresponds to the arbitrary normalization of the Hamiltonian. Notably, this velocity $v$ is identical to the velocity of the bulk CFT Hamiltonian (further discussions see Supplementary Note 2 \cite{SM}).

The state-operator correspondence offers distinct advantages for studying CFTs. Firstly, it provides direct access to information regarding whether the conformal symmetry emerges in the IR. Secondly, it enables an efficient extraction of various conformal data, such as scaling dimensions and OPE coefficients of primaries. The key step involves studying a quantum Hamiltonian on the sphere geometry. However, for 3D CFTs, this was challenging as no regular lattice could fit $S^2$. Recently, this fundamental obstacle was overcome through a scheme called ``fuzzy sphere regularization" \cite{ZHHHH2022}, and its superior capabilities have been convincingly demonstrated \cite{ZHHHH2022,hu2023operator,Han2023Conformal,Zhou2023SO5}. Below we discuss how to adapt the fuzzy sphere regularization scheme to solve dCFTs. We will focus on the case of magnetic line defect of the 3D Ising CFT, but the generalizations to other cases should be straightforward.

\subsection*{Magnetic line defect on the fuzzy sphere}
The fuzzy sphere regularization~\cite{ZHHHH2022} considers a quantum mechanical model describing fermions moving on a sphere with a $4\pi s$ magnetic monopole at the center. The model is generically described by a Hamiltonian $H = H_{\textrm{kin}} + H_{\textrm{int}}$, where $H_{\textrm{kin}}$ represents the kinetic energy of fermions, and its eigenstates form quantized Landau levels described by the monopole Harmonics $Y^{(s)}_{n+s, m}(\theta,\varphi)$~\cite{WuYangmonopole}. Here, $n=0, 1, \cdots$ denotes the Landau level index, and $(\theta, \varphi)$ are the spherical coordinates. We consider the limit where $H_{\textrm{kin}}$ is much larger than the interaction $H_{\textrm{int}}$, allowing us to project the system onto the lowest Landau level (i.e. $n=0$), resulting in a fuzzy sphere~\cite{madore1992fuzzy}.

The 3D Ising transition on the fuzzy sphere can be realized by two-flavor fermions $\bm \psi^\dag = (\psi_\uparrow^\dag, \psi_\downarrow^\dag)$ with interactions that mimic a 2+1D transverse Ising model on the sphere, 
\begin{align}
H_0 = &  \int R^4 d\Omega_a d\Omega_b \,  U(\Omega_{ab}) (n^0(\Omega_a)n^0(\Omega_b) -n^z(\Omega_a)n^z(\Omega_b))\nonumber \\
&- h  \int R^2 d \Omega \, n^x(\Omega).    
\end{align}
Here we are using the spherical coordinate $\Omega=(\theta,\phi)$ and $R$ is the sphere radius. The density operators are defined as  $n^\alpha (\Omega) = \bm \psi^\dag(\Omega) \sigma^a \bm \psi(\Omega)$, where $\sigma^{x,y,z}$ are the Pauli matrices and $\sigma^0$ is the identity matrix. $U(\Omega_{ab})$ encodes the Ising density-density interaction as $U(\Omega_{ab})=\frac{g_0}{R^2}\delta(\Omega_{ab})+\frac{g_1}{R^4}\nabla^2 \delta(\Omega_{ab})$. 
One can tune the transverse field $h$ to realize a phase transition which falls into the 2+1D Ising universality class \cite{ZHHHH2022}. In the following, we set $U(\Omega_{ab})$ and $h$ the same as the bulk Ising CFT that has been identified in \cite{ZHHHH2022}. 

To study the magnetic line defect of 3D Ising CFT, we add $0+1$D point-like magnetic impurities located at sphere's north and south pole, modeled by a Hamiltonian term, 
\begin{equation}
H_{d}  = 2\pi h_d (n^z(\theta=0,\varphi=0)) + n^z(\theta=\pi,\varphi=0) ), 
\end{equation}
where $h_d$ controls the strength of the magnetic impurities. This type of defect can be artificially realized in experiments \cite{LAW2001159,Fisher_Gennes}. 
Crucially, the defect term $H_d$ breaks the Ising $\mathbb{Z}_2$ symmetry, causing the $\sigma$ field (of the 3D Ising CFT) to be turned on at the defect. This $\sigma$ deformation is relevant on the line defect ($\Delta_\sigma \approx 0.518 < 1$~\cite{RMP_CB}), driving the system to flow to a nontrivial fixed point, conjectured to be a conformal defect. This fixed point is expected to be an attractive fixed point~\cite{Andrea2014,Hanke2000,Cuomo2022magneticDefect}, implying that regardless of the strength of $h_d$, the defect will flow to the same conformal defect fixed point (see Fig. \ref{fig:displacement_flow}(a)). Next we will provide compelling numerical evidence to support this conjecture.


\subsection*{Emergent conformal symmetry and operator spectrum}
The energy spectrum of the defect Hamiltonian ($H_0+H_d$) is expected to be proportional to the defect operators' scaling dimensions, up to a non-universal velocity in Eq.~\eqref{eq:correspondence}. Here we determine the velocity  using the bulk CFT Hamiltonian ($H_0$) by setting the $\sigma$ state to have $\Delta_\sigma = 0.518149$ \cite{RMP_CB}. The defect term $H_d$ breaks the sphere rotation $SO(3)$ down to $SO(2)$, so each eigenstate has a well defined $SO(2)$ quantum number $L_z$. Akin to the stress tensor of the bulk CFT, there exists a special primary operator in dCFT due to the broken of translation symmetry, dubbed the displacement operator $\hat D$ \cite{Billo2013,Billo2016defect,MCAVITY1995}, which has $L_z=\pm 1$ and a protected scaling dimension $\Delta_{\hat D}=2$. Fig. \ref{fig:displacement_flow}(b-c)  depicts $\Delta_{\hat D}$ via the state-operator correspondence (Eq.  \eqref{eq:correspondence}) for various defect strength $h_d$ and system sizes. It clearly shows that the obtained $\Delta_{\hat D}$ are very close to $2$, for different defect strengths $h_d$, which indicates an attractive conformal fixed point at $h_d=\infty$ (see Supplementary Note 3 and 6\cite{SM}).
In what follows, we present the representative results for $h_d=300$ and we ensure the conclusions are insensitive to the choice of $h_d$.

\begin{table}
\setlength{\tabcolsep}{0.2cm}
\renewcommand{\arraystretch}{1.4}
    \centering
    \caption{
    Scaling dimensions of primary operators in the magnetic line defect of 3D Ising CFT, determined through the state-operator correspondence on the fuzzy sphere. 
    Please see a detailed analysis of errors and finite-size extrapolation in Supplementary Note 2-3 \cite{SM}.
}\label{tab:primary}
\begin{tabular}{ccc|ccc} 
\hline\hline
\multicolumn{3}{c|}{$L_z=0$}& \multicolumn{3}{c}{$L_z=1$} \\
    \hline
$\hat \phi$ & $\hat \phi'$ & $\hat \phi''$  & $\hat D$ & $\hat \phi_1$ & $\hat \phi_1'$\\    
 1.63(6) & 3.12(10) & 4.06(18) & 2.05(7) & 3.58(7) &  4.64(14)  \\   
 \hline\hline
\end{tabular} 
\end{table}

\begin{figure}[b]
\includegraphics[width=0.99\linewidth]{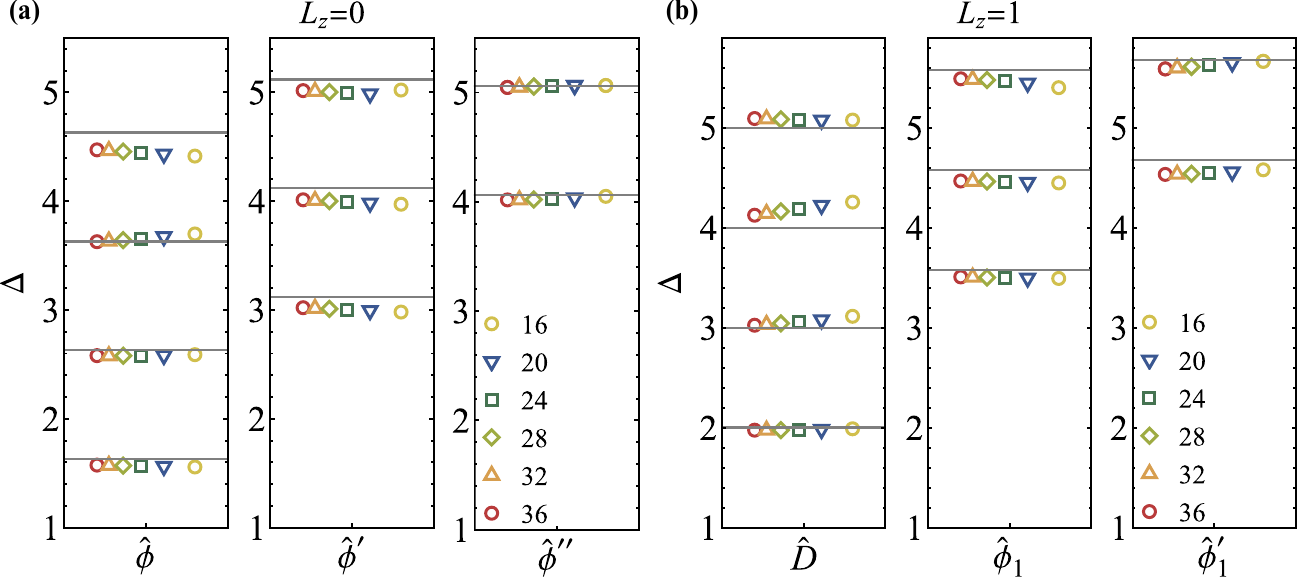}
\caption{\textbf{Conformal tower of defect primaries}. Defect primary fields and their descendants with global symmetry (a) $L_z=0$ and (b) $L_z=1$. 
The grey horizon lines stand for extrapolated values for primaries and their integer-spaced descendants.
Different colored symbols represent the results based on various system sizes. By increasing system size $N$ all of the scaling dimensions approach the theoretical values consistently, supporting an emergent conformal symmetry in the thermodynamic limit.
\label{fig:tower}
}
\end{figure}

We further establish the emergent conformal symmetry by confirming that the excitation spectra form representations of $SO(2,1)$. The generators of $SO(2,1)$ are the dilation $D$, translation $P$, and special conformal transformation $K$. It is important to note that $P$ and $K$ do not have any Lorentz index due to the triviality of the Lorentz symmetry, i.e., $SO(1)$.
For each primary operator, we have descendants generated by the translation, $P^n \hat{O}$, whose scaling dimension is $\Delta_{P^n \hat{O}}=\Delta_{\hat{O}}+n$, and its $SO(2)$ quantum number $L_z$ remains unchanged. Fig. \ref{fig:tower} displays our numerical data of the low-lying energy spectrum, clearly exhibiting the emergent conformal symmetry, i.e., approximate integer spacing between each primary and its descendants.
These observations firmly establish that the magnetic line defect of the 3D Ising CFT flows to a conformal defect with a conformal symmetry of $SO(2,1)$.

From our numerical data, we can identify five low-lying defect primary operators in addition to $\hat{D}$, as listed in Tab. \ref{tab:primary}. Notably, all these operators are found to be irrelevant (i.e., $\Delta>1$), which is consistent with the observation of an attractive defect fixed point.
Our lowest-lying operator $\hat{\phi}$ has $L_z=0$ and $\Delta_{\hat{\phi}}\approx 1.63(6)$. This value is in good agreement with Monte Carlo simulations, e.g. $1.60(5)$~\cite{Assaad2017}, $1.52(6)$~\cite{Assaad2017}, and $1.40(3)$~\cite{Allais2014}, as well as with the perturbative $\varepsilon$-expansion computation of $\sim1.55(14)$ Ref.~\cite{Cuomo2022magneticDefect}.
The second low-lying operator in the $L_z=0$ sector has $\Delta_{\hat{\phi'}} = 3.12(10)$, which significantly deviates from the $\varepsilon$-expansion value of $\Delta\approx 4.33+O(\varepsilon^2)$ (it was called $\hat s_+$ in ~\cite{Cuomo2022magneticDefect}). This suggests a large sub-leading correction in the $\varepsilon$-expansion. All other primary operators identified in our study
have not been computed by any other methods.
It is essential to mention that the scaling dimensions in Tab. \ref{tab:primary} are obtained by the finite-size extrapolating (see details in Supplementary Note 2 \cite{SM}), and the data at finite $N$ is already very close to the extrapolated value (The finite-size extrapolation improve the results by around $2\%$). One can also improve the accuracy by making use of conformal perturbation~\cite{Rychkov2023Icosahedron}.

\begin{figure}[t]
\includegraphics[width=0.75\linewidth]{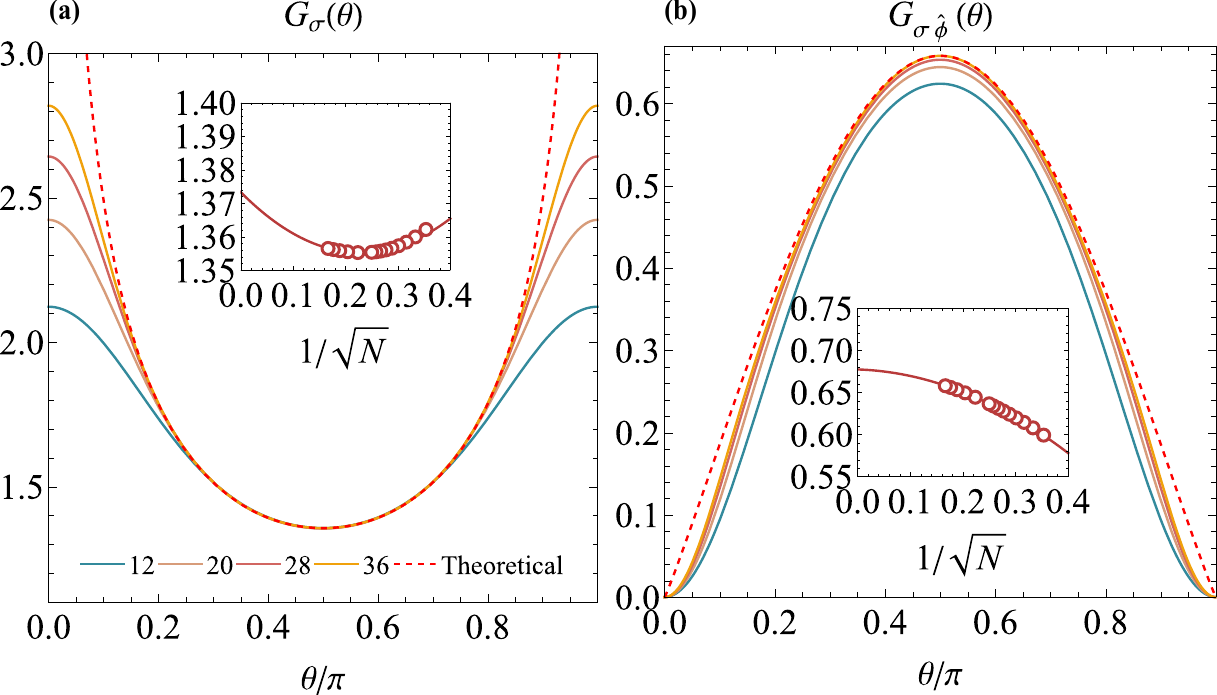}
\caption{\textbf{Correlators involving defect}. The angle dependence of (a) correlator $G_{\sigma}(\theta)$ and (b) $G_{\sigma\hat{\phi}}(\theta)$, for system sizes ranging from $N=12$ to $36$. The dashed lines correspond to  theoretical correlator in Eq.~\eqref{eq:sphere_corr} with $b_{\sigma \hat O_2}$ and $\Delta_{\hat O_2}$ from the $N=36$ curves.
The insets are finite-size scaling analysis by setting $\theta=\pi/2$, respectively giving one-point OPE coefficient $a_\sigma\approx 1.37(1)$ and bulk-defect OPE coefficient $b_{\sigma\hat \phi}\approx 0.68(1)$. }
\label{fig:sigma}
\end{figure}

Additionally, to verify the physics of dCFT presented here is independent of the specific value  $h_d$, we directly study the spectrum at $h_d=\infty$ (see Supplementary Note 6 \cite{SM}). Comparing it with the results at $h_d=300$, we found them to be consistent with each other. This result indicates that the large $h_d$ regime shares the same dCFT and also supports that the fixed point of dCFT indeed resides at $h_d=\infty$.

\subsection*{Correlators and OPE coefficients} Using Weyl transformation, we can map the bulk-defect correlators in Eq.~\eqref{eq:1pt}, Eq.~\eqref{eq:2pt} in $\mathbb R^3$ to the correlators on cylinder $S^2\times \mathbb R$ (see Supplementary Note 1 \cite{SM}),  
\begin{equation}\label{eq:sphere_corr}
   G_{O_1\hat{O}_2}\equiv \frac{ \langle \hat{ 1} | O_1(\tau=0, \theta) |\hat O_2\rangle}{\langle 1| O_1(\tau=0, \theta) |O_{1}\rangle} = \frac{b_{O_1 \hat O_2}}{(\sin \theta)^{\Delta_1-\hat \Delta_2}}.
\end{equation} 
The bulk operator $O_1$ is positioned at a point that has an angle $\theta$ with respect to the north pole. In the denominator, we use the states of the bulk CFT, while in the numerator, we use the states of the dCFT. The one-point bulk correlator corresponds to taking $|\hat O_2\rangle$ to be the ground state of the defect, i.e., $| \hat{1} \rangle$.

In the fuzzy sphere model, we can use the spin operators $n^z$ and $n^x$ to approximate the bulk CFT primary operators $\sigma$ and $\epsilon$~\cite{hu2023operator,Han2023Conformal}. For example, the correlator between the bulk primary $\sigma$ and a defect primary operator $\hat O_2$ is computed by, 
\begin{equation}\label{eq:fuzzy_corr}
G_{\sigma\hat{O}_2}\equiv \frac{ \langle \hat{1} | n^z(\tau=0, \theta) |\hat O_2\rangle}{\langle 1| n^z(\tau=0, \theta) |\sigma\rangle} = \frac{b_{\sigma \hat O_2}}{(\sin \theta)^{\Delta_\sigma-\hat \Delta_2}} + O(N^{-1/2}). 
\end{equation}
Here, $\Delta_\sigma \approx 0.518149$, and the first-order correction $O(N^{-1/2})$ comes from the descendant operator $\partial_\mu \sigma$ contained in $n^z$. Fig.~\ref{fig:sigma} illustrates the one-point bulk correlator $G_{\sigma}(\theta)$ and bulk-defect correlator $G_{\sigma \hat \phi}(\theta)$ for different system sizes $N=12-36$.
Both correlators agree perfectly with the CFT prediction Eq.\eqref{eq:sphere_corr}, except for the small $\theta$ regime. It is worth noting that the one-point correlator $G_{\sigma}(\theta)$ is divergent at $\theta=0, \pi$ and reaches a minimum at $\theta=\pi/2$. In contrast, the bulk-defect correlator $G_{\sigma \hat \phi}(\theta)$ has an opposite behavior (because $\Delta_\sigma-\Delta_{\hat \phi}<0$); it vanishes at $\theta=0, \pi$ and reaches a maximum at $\theta=\pi/2$. These behaviors are nicely reproduced in our data, which is highly nontrivial because computationally the only difference for the two correlators is the choice of $|\hat O_2\rangle$ in Eq.~\eqref{eq:fuzzy_corr}.

\begin{table}
\setlength{\tabcolsep}{0.1cm}
\renewcommand{\arraystretch}{1.4}
    \centering
    \caption{ Bulk-to-defect OPE coefficients magnetic line defect of 3D Ising CFT. $C_{\hat D}$ is computed by Eq.~\eqref{eq:CD} using $\sigma$ and $\epsilon$. }
    \label{tab:OPE}
\begin{tabular}{cccccc} 
\hline\hline
 $a_{\sigma}$ &$b_{\sigma \hat \phi}$  & $a_{\epsilon}$ & $b_{\epsilon \hat \phi}$ & $C_{\hat D}$ by $\sigma$ & $C_{\hat D}$  by $\epsilon$ \\
 1.37(1) & 0.68(1)  &  1.31(19) & 1.63(4) & 0.27(1) & 0.30(8)\\   
\hline 
\hline 
\end{tabular} 
\end{table}

We can further extract the bulk-defect OPE coefficients from $G_{O_1\hat O_2}(\theta=\pi/2) = b_{O_1\hat O_2}$, and the results are summarized in Tab.~\ref{tab:OPE}. None of these OPE coefficients was computed non-perturbatively before. There are perturbative computations for  $a_\sigma$ and $a_\epsilon$~\cite{Cuomo2022magneticDefect} from $\varepsilon$ expansion, giving $a_\sigma^2\approx 3.476+O(\varepsilon^2)$ (i.e. $a_\sigma\approx 1.86$) and $a_\epsilon \approx 1.83 + O(\varepsilon^2)$. Our estimates are $a_\sigma = 1.37(1)$ and $a_\epsilon=1.31(19)$, it will be interesting to compute higher order corrections in the $\varepsilon$-expansion. Moreover, using the Ward identity of any bulk operator ($O$)~\cite{Billo2016defect}, we can extract Zamolodchikov norm
\begin{equation}\label{eq:CD}
\sqrt{2C_{\hat D}} = \frac{2}{\pi} \frac{\Delta_O G_{O}(\theta=\pi/2)}{ G_{O\hat D}(\theta=\pi/2) } .   
\end{equation}
The estimates using $\sigma$ and $\epsilon$ gives $C_{\hat D} = 0.27(1)$ and $C_{\hat D} = 0.30(8)$, respectively.

\section*{Summary and discussion}
We have outlined a systematic procedure to study defect conformal field theory (dCFT) using the recently-proposed fuzzy sphere regularization scheme. As a concrete application, we investigated the magnetic line defect of 3D Ising CFT and provided clear evidence that it flows to a conformal defect. Crucially, we accurately computed a number of conformal data of this dCFT, including defect primaries' scaling dimensions and bulk-defect OPE coefficients. 
As far as we know, most of the conformal data of dCFT reported here have never been studied in a microscopic model before, thus this conformal information paves the way for exploring the rich physics in 3D Ising CFTs.

Looking forward, the current setup can be readily applied to the study of various types of defects in distinct 3D CFTs, potentially resolving numerous open questions and offering insights into defects in CFTs.
For example, the plane defect ($p=2$ in Eq. \ref{eq:Ham_dCFT}), which may resemble surface critical phenomena, is interesting to investigate. It is also highly desired to study 3D dCFTs in a broad universality class (e.g. Wilson-Fisher $O(N)$ critical point).
Moreover, the results of current work provide a necessary input in the study of infrared data for the dCFT within the numerical conformal bootstrap \cite{Gaiotto2014,Liendo2013,RMP_CB}.
Taking it further would potentially be very interesting to study the dCFT in holography and string theory.


\section*{Methods}
The model $H_0+H_d$ for the magnetic line defect of 3D Ising CFT  is a continuous model with fully local interaction in the spatial space. 
In practice, we consider the second quantization form of this model by the projecting $H_0+H_d$ to the lowest Landau level (fuzzy sphere), 
using $\psi_a (\Omega) = \frac{1}{\sqrt{N}} \sum_{m=-s}^s c_{m, a}  Y^{(s)}_{s, m} (\Omega)$ (we are using a slightly different convention compared to Ref.~\cite{ZHHHH2022}). 
Here $N=2s+1$ playing the role of system size $N\sim R^2$, and we simply replace $R^2$ with $N$ during the projection. This lowest Landau level projection leads to a second quantized Hamiltonian defined by fermionic operators $c_{m,a}$, and similar models have been extensively studied in the context of the quantum Hall effect~\cite{Sphere_LL_Haldane}. Numerically, this model can be simulated using various techniques such as exact diagonalization and density matrix renormalization group (DMRG) \cite{SWhite1992,Feiguin2008}. We perform DMRG calculations with bond dimensions up to $D=5000$, and for the largest system size $N=36$, the maximum truncation errors for the ground state and the tenth excited state are $1.37 \times 10^{-9}$ and $1.96 \times 10^{-8}$, respectively. We explicitly impose two $U(1)$ symmetries, i.e., fermion number and $SO(2)$ angular momentum.

\noindent {\bf Data availability}\\
All data are included in this published article and Supplementary Information files.

\noindent {\bf Code availability}\\
The codes used to generate data and plots are available from the corresponding
author upon request. The DMRG data are generated using the software “ITensor 3(C++ Version)”\cite{fishman2020itensor}.

\noindent {\bf Acknowledgments}\\
We thank Davide Gaiotto for stimulating discussions that initiated this project. 
LDH and WZ were supported by National Natural Science Foundation of China (No.~92165102, 11974288) and National key R$\&$D program (No. 2022YFA1402204).  
Research at Perimeter Institute is supported in part by the Government of Canada through the Department of Innovation, Science and Industry Canada and by the Province of Ontario through the Ministry of Colleges and Universities. YCH thanks the hospitality of Bootstrap 2023 at  ICTP South American Institute for Fundamental Research, where part of this project was done.\\

\noindent {\bf Author contributions}\\
W.Z.\ and Y.-C.H.\ initiated the project, L.D.H. performed the simulations.
All authors contributed equally to the analysis of the data and writing of the manuscript.\\

\noindent {\bf Competing interests}\\
The authors declare no competing interests.

\noindent {\bf Tables}\\
TABLE. I. Scaling dimensions of primary operators in the magnetic line defect of 3D Ising CFT.\\
TABLE. II. Bulk-to-defect OPE coefficients magnetic line defect of 3D Ising CFT.

\noindent {\bf Figure capations}\\
FIGURE. 1. Schematic plot of the defect in 3D.\\
FIGURE. 2. Defect-induced attractive fixed point.\\
FIGURE. 3. Conformal tower of defect primaries.\\
FIGURE. 4. Correlators involving defect.
\bibliography{cft_defect}

\begin{thebibliography}{62}%
\makeatletter
\providecommand \@ifxundefined [1]{%
 \@ifx{#1\undefined}
}%
\providecommand \@ifnum [1]{%
 \ifnum #1\expandafter \@firstoftwo
 \else \expandafter \@secondoftwo
 \fi
}%
\providecommand \@ifx [1]{%
 \ifx #1\expandafter \@firstoftwo
 \else \expandafter \@secondoftwo
 \fi
}%
\providecommand \natexlab [1]{#1}%
\providecommand \enquote  [1]{``#1''}%
\providecommand \bibnamefont  [1]{#1}%
\providecommand \bibfnamefont [1]{#1}%
\providecommand \citenamefont [1]{#1}%
\providecommand \href@noop [0]{\@secondoftwo}%
\providecommand \href [0]{\begingroup \@sanitize@url \@href}%
\providecommand \@href[1]{\@@startlink{#1}\@@href}%
\providecommand \@@href[1]{\endgroup#1\@@endlink}%
\providecommand \@sanitize@url [0]{\catcode `\\12\catcode `\$12\catcode
  `\&12\catcode `\#12\catcode `\^12\catcode `\_12\catcode `\%12\relax}%
\providecommand \@@startlink[1]{}%
\providecommand \@@endlink[0]{}%
\providecommand \url  [0]{\begingroup\@sanitize@url \@url }%
\providecommand \@url [1]{\endgroup\@href {#1}{\urlprefix }}%
\providecommand \urlprefix  [0]{URL }%
\providecommand \Eprint [0]{\href }%
\providecommand \doibase [0]{http://dx.doi.org/}%
\providecommand \selectlanguage [0]{\@gobble}%
\providecommand \bibinfo  [0]{\@secondoftwo}%
\providecommand \bibfield  [0]{\@secondoftwo}%
\providecommand \translation [1]{[#1]}%
\providecommand \BibitemOpen [0]{}%
\providecommand \bibitemStop [0]{}%
\providecommand \bibitemNoStop [0]{.\EOS\space}%
\providecommand \EOS [0]{\spacefactor3000\relax}%
\providecommand \BibitemShut  [1]{\csname bibitem#1\endcsname}%
\let\auto@bib@innerbib\@empty
\bibitem [{\citenamefont {Wilson}(1975)}]{Wilson1975}%
  \BibitemOpen
  \bibfield  {author} {\bibinfo {author} {\bibfnamefont {Kenneth~G.}\
  \bibnamefont {Wilson}},\ }\bibfield  {title} {\enquote {\bibinfo {title} {The
  renormalization group: Critical phenomena and the kondo problem},}\ }\href
  {\doibase 10.1103/RevModPhys.47.773} {\bibfield  {journal} {\bibinfo
  {journal} {Rev. Mod. Phys.}\ }\textbf {\bibinfo {volume} {47}},\ \bibinfo
  {pages} {773--840} (\bibinfo {year} {1975})}\BibitemShut {NoStop}%
\bibitem [{\citenamefont {Nayak}\ \emph {et~al.}(2008)\citenamefont {Nayak},
  \citenamefont {Simon}, \citenamefont {Stern}, \citenamefont {Freedman},\ and\
  \citenamefont {Das~Sarma}}]{anyon_RMP_2008}%
  \BibitemOpen
  \bibfield  {author} {\bibinfo {author} {\bibfnamefont {Chetan}\ \bibnamefont
  {Nayak}}, \bibinfo {author} {\bibfnamefont {Steven~H.}\ \bibnamefont
  {Simon}}, \bibinfo {author} {\bibfnamefont {Ady}\ \bibnamefont {Stern}},
  \bibinfo {author} {\bibfnamefont {Michael}\ \bibnamefont {Freedman}}, \ and\
  \bibinfo {author} {\bibfnamefont {Sankar}\ \bibnamefont {Das~Sarma}},\
  }\bibfield  {title} {\enquote {\bibinfo {title} {Non-abelian anyons and
  topological quantum computation},}\ }\href {\doibase
  10.1103/RevModPhys.80.1083} {\bibfield  {journal} {\bibinfo  {journal} {Rev.
  Mod. Phys.}\ }\textbf {\bibinfo {volume} {80}},\ \bibinfo {pages}
  {1083--1159} (\bibinfo {year} {2008})}\BibitemShut {NoStop}%
\bibitem [{\citenamefont {Hasan}\ and\ \citenamefont
  {Kane}(2010)}]{Hasan2010Review}%
  \BibitemOpen
  \bibfield  {author} {\bibinfo {author} {\bibfnamefont {M.~Z.}\ \bibnamefont
  {Hasan}}\ and\ \bibinfo {author} {\bibfnamefont {C.~L.}\ \bibnamefont
  {Kane}},\ }\bibfield  {title} {\enquote {\bibinfo {title} {Colloquium:
  Topological insulators},}\ }\href {\doibase 10.1103/RevModPhys.82.3045}
  {\bibfield  {journal} {\bibinfo  {journal} {Rev. Mod. Phys.}\ }\textbf
  {\bibinfo {volume} {82}},\ \bibinfo {pages} {3045--3067} (\bibinfo {year}
  {2010})}\BibitemShut {NoStop}%
\bibitem [{\citenamefont {Qi}\ and\ \citenamefont
  {Zhang}(2011)}]{Qi2011Reivew}%
  \BibitemOpen
  \bibfield  {author} {\bibinfo {author} {\bibfnamefont {Xiao-Liang}\
  \bibnamefont {Qi}}\ and\ \bibinfo {author} {\bibfnamefont {Shou-Cheng}\
  \bibnamefont {Zhang}},\ }\bibfield  {title} {\enquote {\bibinfo {title}
  {Topological insulators and superconductors},}\ }\href {\doibase
  10.1103/RevModPhys.83.1057} {\bibfield  {journal} {\bibinfo  {journal} {Rev.
  Mod. Phys.}\ }\textbf {\bibinfo {volume} {83}},\ \bibinfo {pages}
  {1057--1110} (\bibinfo {year} {2011})}\BibitemShut {NoStop}%
\bibitem [{\citenamefont {Wilson}(1974)}]{Wilson1974Confinement}%
  \BibitemOpen
  \bibfield  {author} {\bibinfo {author} {\bibfnamefont {Kenneth~G.}\
  \bibnamefont {Wilson}},\ }\bibfield  {title} {\enquote {\bibinfo {title}
  {Confinement of quarks},}\ }\href {\doibase 10.1103/PhysRevD.10.2445}
  {\bibfield  {journal} {\bibinfo  {journal} {Phys. Rev. D}\ }\textbf {\bibinfo
  {volume} {10}},\ \bibinfo {pages} {2445--2459} (\bibinfo {year}
  {1974})}\BibitemShut {NoStop}%
\bibitem [{\citenamefont {Hooft}(1978)}]{hooft1978phase}%
  \BibitemOpen
  \bibfield  {author} {\bibinfo {author} {\bibfnamefont {Gerard’t}\
  \bibnamefont {Hooft}},\ }\bibfield  {title} {\enquote {\bibinfo {title} {On
  the phase transition towards permanent quark confinement},}\ }\href@noop {}
  {\bibfield  {journal} {\bibinfo  {journal} {Nuclear Physics: B}\ }\textbf
  {\bibinfo {volume} {138}},\ \bibinfo {pages} {1--25} (\bibinfo {year}
  {1978})}\BibitemShut {NoStop}%
\bibitem [{\citenamefont {Maldacena}(1999)}]{Maldacena_AdSCFT}%
  \BibitemOpen
  \bibfield  {author} {\bibinfo {author} {\bibfnamefont {Juan}\ \bibnamefont
  {Maldacena}},\ }\bibfield  {title} {\enquote {\bibinfo {title} {The large-n
  limit of superconformal field theories and supergravity},}\ }\href {\doibase
  10.1023/A:1026654312961} {\bibfield  {journal} {\bibinfo  {journal}
  {International Journal of Theoretical Physics}\ }\textbf {\bibinfo {volume}
  {38}},\ \bibinfo {pages} {1113--1133} (\bibinfo {year} {1999})}\BibitemShut
  {NoStop}%
\bibitem [{\citenamefont {Holzhey}\ \emph {et~al.}(1994)\citenamefont
  {Holzhey}, \citenamefont {Larsen},\ and\ \citenamefont
  {Wilczek}}]{HOLZHEY1994}%
  \BibitemOpen
  \bibfield  {author} {\bibinfo {author} {\bibfnamefont {Christoph}\
  \bibnamefont {Holzhey}}, \bibinfo {author} {\bibfnamefont {Finn}\
  \bibnamefont {Larsen}}, \ and\ \bibinfo {author} {\bibfnamefont {Frank}\
  \bibnamefont {Wilczek}},\ }\bibfield  {title} {\enquote {\bibinfo {title}
  {Geometric and renormalized entropy in conformal field theory},}\ }\href
  {\doibase 10.1016/0550-3213(94)90402-2} {\bibfield  {journal} {\bibinfo
  {journal} {Nuclear Physics B}\ }\textbf {\bibinfo {volume} {424}},\ \bibinfo
  {pages} {443 -- 467} (\bibinfo {year} {1994})}\BibitemShut {NoStop}%
\bibitem [{\citenamefont {Calabrese}\ and\ \citenamefont
  {Cardy}(2004)}]{calabrese2004entanglement}%
  \BibitemOpen
  \bibfield  {author} {\bibinfo {author} {\bibfnamefont {Pasquale}\
  \bibnamefont {Calabrese}}\ and\ \bibinfo {author} {\bibfnamefont {John}\
  \bibnamefont {Cardy}},\ }\bibfield  {title} {\enquote {\bibinfo {title}
  {Entanglement entropy and quantum field theory},}\ }\href {\doibase
  10.1088/1742-5468/2004/06/P06002} {\bibfield  {journal} {\bibinfo  {journal}
  {Journal of Statistical Mechanics: Theory and Experiment}\ }\textbf {\bibinfo
  {volume} {2004}},\ \bibinfo {pages} {P06002} (\bibinfo {year}
  {2004})}\BibitemShut {NoStop}%
\bibitem [{\citenamefont {Philippe~Francesco}(1997)}]{yellowbook}%
  \BibitemOpen
  \bibfield  {author} {\bibinfo {author} {\bibfnamefont {David~Sénéchal}\
  \bibnamefont {Philippe~Francesco}, \bibfnamefont {Pierre~Mathieu}},\
  }\href@noop {} {\emph {\bibinfo {title} {Conformal Field Theory}}},\ Graduate
  Texts in Contemporary Physics\ (\bibinfo  {publisher} {Springer New York,
  NY},\ \bibinfo {year} {1997})\BibitemShut {NoStop}%
\bibitem [{\citenamefont {Cardy}(1996)}]{Cardy_book}%
  \BibitemOpen
  \bibfield  {author} {\bibinfo {author} {\bibfnamefont {J.}~\bibnamefont
  {Cardy}},\ }\href@noop {} {\emph {\bibinfo {title} {Scaling and
  Renormalization in Statistical Physics}}}\ (\bibinfo  {publisher} {Cambridge
  University Press, Cambridge, England},\ \bibinfo {year} {1996})\BibitemShut
  {NoStop}%
\bibitem [{\citenamefont {Cardy}(1984{\natexlab{a}})}]{CARDY1984surface}%
  \BibitemOpen
  \bibfield  {author} {\bibinfo {author} {\bibfnamefont {John~L.}\ \bibnamefont
  {Cardy}},\ }\bibfield  {title} {\enquote {\bibinfo {title} {Conformal
  invariance and surface critical behavior},}\ }\href {\doibase
  https://doi.org/10.1016/0550-3213(84)90241-4} {\bibfield  {journal} {\bibinfo
   {journal} {Nuclear Physics B}\ }\textbf {\bibinfo {volume} {240}},\ \bibinfo
  {pages} {514--532} (\bibinfo {year} {1984}{\natexlab{a}})}\BibitemShut
  {NoStop}%
\bibitem [{\citenamefont {Cardy}(1989)}]{cardy1989boundary}%
  \BibitemOpen
  \bibfield  {author} {\bibinfo {author} {\bibfnamefont {John~L.}\ \bibnamefont
  {Cardy}},\ }\bibfield  {title} {\enquote {\bibinfo {title} {Boundary
  conditions, fusion rules and the verlinde formula},}\ }\href {\doibase
  https://doi.org/10.1016/0550-3213(89)90521-X} {\bibfield  {journal} {\bibinfo
   {journal} {Nuclear Physics B}\ }\textbf {\bibinfo {volume} {324}},\ \bibinfo
  {pages} {581--596} (\bibinfo {year} {1989})}\BibitemShut {NoStop}%
\bibitem [{\citenamefont {Cardy}\ and\ \citenamefont
  {Lewellen}(1991)}]{cardy1991bulkboundary}%
  \BibitemOpen
  \bibfield  {author} {\bibinfo {author} {\bibfnamefont {John~L.}\ \bibnamefont
  {Cardy}}\ and\ \bibinfo {author} {\bibfnamefont {David~C.}\ \bibnamefont
  {Lewellen}},\ }\bibfield  {title} {\enquote {\bibinfo {title} {Bulk and
  boundary operators in conformal field theory},}\ }\href {\doibase
  https://doi.org/10.1016/0370-2693(91)90828-E} {\bibfield  {journal} {\bibinfo
   {journal} {Physics Letters B}\ }\textbf {\bibinfo {volume} {259}},\ \bibinfo
  {pages} {274--278} (\bibinfo {year} {1991})}\BibitemShut {NoStop}%
\bibitem [{\citenamefont {Diehl}\ and\ \citenamefont
  {Dietrich}(1981)}]{Diehl1981Field-theoretical}%
  \BibitemOpen
  \bibfield  {author} {\bibinfo {author} {\bibfnamefont {H.~W.}\ \bibnamefont
  {Diehl}}\ and\ \bibinfo {author} {\bibfnamefont {S.}~\bibnamefont
  {Dietrich}},\ }\bibfield  {title} {\enquote {\bibinfo {title}
  {Field-theoretical approach to multicritical behavior near free surfaces},}\
  }\href {\doibase 10.1103/PhysRevB.24.2878} {\bibfield  {journal} {\bibinfo
  {journal} {Phys. Rev. B}\ }\textbf {\bibinfo {volume} {24}},\ \bibinfo
  {pages} {2878--2880} (\bibinfo {year} {1981})}\BibitemShut {NoStop}%
\bibitem [{\citenamefont {{McAvity}}\ and\ \citenamefont
  {{Osborn}}(1993)}]{McAvity1993EMT}%
  \BibitemOpen
  \bibfield  {author} {\bibinfo {author} {\bibfnamefont {D.~M.}\ \bibnamefont
  {{McAvity}}}\ and\ \bibinfo {author} {\bibfnamefont {H.}~\bibnamefont
  {{Osborn}}},\ }\bibfield  {title} {\enquote {\bibinfo {title}
  {{Energy-momentum tensor in conformal field theories near a boundary}},}\
  }\href {\doibase 10.1016/0550-3213(93)90005-A} {\bibfield  {journal}
  {\bibinfo  {journal} {Nuclear Physics B}\ }\textbf {\bibinfo {volume}
  {406}},\ \bibinfo {pages} {655--680} (\bibinfo {year} {1993})},\ \Eprint
  {http://arxiv.org/abs/hep-th/9302068} {arXiv:hep-th/9302068 [hep-th]}
  \BibitemShut {NoStop}%
\bibitem [{\citenamefont {McAvity}\ and\ \citenamefont
  {Osborn}(1995)}]{MCAVITY1995}%
  \BibitemOpen
  \bibfield  {author} {\bibinfo {author} {\bibfnamefont {D.M.}\ \bibnamefont
  {McAvity}}\ and\ \bibinfo {author} {\bibfnamefont {H.}~\bibnamefont
  {Osborn}},\ }\bibfield  {title} {\enquote {\bibinfo {title} {Conformal field
  theories near a boundary in general dimensions},}\ }\href {\doibase
  https://doi.org/10.1016/0550-3213(95)00476-9} {\bibfield  {journal} {\bibinfo
   {journal} {Nuclear Physics B}\ }\textbf {\bibinfo {volume} {455}},\ \bibinfo
  {pages} {522--576} (\bibinfo {year} {1995})}\BibitemShut {NoStop}%
\bibitem [{\citenamefont {{Bill{\'o}}}\ \emph {et~al.}(2013)\citenamefont
  {{Bill{\'o}}}, \citenamefont {{Caselle}}, \citenamefont {{Gaiotto}},
  \citenamefont {{Gliozzi}}, \citenamefont {{Meineri}},\ and\ \citenamefont
  {{Pellegrini}}}]{Billo2013}%
  \BibitemOpen
  \bibfield  {author} {\bibinfo {author} {\bibfnamefont {M.}~\bibnamefont
  {{Bill{\'o}}}}, \bibinfo {author} {\bibfnamefont {M.}~\bibnamefont
  {{Caselle}}}, \bibinfo {author} {\bibfnamefont {D.}~\bibnamefont
  {{Gaiotto}}}, \bibinfo {author} {\bibfnamefont {F.}~\bibnamefont
  {{Gliozzi}}}, \bibinfo {author} {\bibfnamefont {M.}~\bibnamefont
  {{Meineri}}}, \ and\ \bibinfo {author} {\bibfnamefont {R.}~\bibnamefont
  {{Pellegrini}}},\ }\bibfield  {title} {\enquote {\bibinfo {title} {{Line
  defects in the 3d Ising model}},}\ }\href {\doibase
  https://doi.org/10.1007/JHEP07(2013)055} {\bibfield  {journal} {\bibinfo
  {journal} {Journal of High Energy Physics}\ }\textbf {\bibinfo {volume}
  {2013}},\ \bibinfo {eid} {55} (\bibinfo {year} {2013})},\ \Eprint
  {http://arxiv.org/abs/1304.4110} {arXiv:1304.4110 [hep-th]} \BibitemShut
  {NoStop}%
\bibitem [{\citenamefont {{Bill{\`o}}}\ \emph {et~al.}(2016)\citenamefont
  {{Bill{\`o}}}, \citenamefont {{Gon{\c{c}}alves}}, \citenamefont {{Lauria}},\
  and\ \citenamefont {{Meineri}}}]{Billo2016defect}%
  \BibitemOpen
  \bibfield  {author} {\bibinfo {author} {\bibfnamefont {Marco}\ \bibnamefont
  {{Bill{\`o}}}}, \bibinfo {author} {\bibfnamefont {Vasco}\ \bibnamefont
  {{Gon{\c{c}}alves}}}, \bibinfo {author} {\bibfnamefont {Edoardo}\
  \bibnamefont {{Lauria}}}, \ and\ \bibinfo {author} {\bibfnamefont {Marco}\
  \bibnamefont {{Meineri}}},\ }\bibfield  {title} {\enquote {\bibinfo {title}
  {{Defects in conformal field theory}},}\ }\href {\doibase
  10.1007/JHEP04(2016)091} {\bibfield  {journal} {\bibinfo  {journal} {Journal
  of High Energy Physics}\ }\textbf {\bibinfo {volume} {2016}},\ \bibinfo {eid}
  {91} (\bibinfo {year} {2016})},\ \Eprint {http://arxiv.org/abs/1601.02883}
  {arXiv:1601.02883 [hep-th]} \BibitemShut {NoStop}%
\bibitem [{\citenamefont {Liendo}\ \emph {et~al.}(2013)\citenamefont {Liendo},
  \citenamefont {Rastelli},\ and\ \citenamefont {van Rees}}]{Liendo2013}%
  \BibitemOpen
  \bibfield  {author} {\bibinfo {author} {\bibfnamefont {P.}~\bibnamefont
  {Liendo}}, \bibinfo {author} {\bibfnamefont {L.}~\bibnamefont {Rastelli}}, \
  and\ \bibinfo {author} {\bibfnamefont {B.C.}\ \bibnamefont {van Rees}},\
  }\bibfield  {title} {\enquote {\bibinfo {title} {{The bootstrap program for
  boundary CFT$_d$.}}}\ }\href {\doibase
  https://doi.org/10.1007/JHEP07(2013)113} {\bibfield  {journal} {\bibinfo
  {journal} {Journal of High Energy Physics}\ }\textbf {\bibinfo {volume}
  {2013}},\ \bibinfo {eid} {113} (\bibinfo {year} {2013})}\BibitemShut
  {NoStop}%
\bibitem [{\citenamefont {Poland}\ \emph {et~al.}(2019)\citenamefont {Poland},
  \citenamefont {Rychkov},\ and\ \citenamefont {Vichi}}]{RMP_CB}%
  \BibitemOpen
  \bibfield  {author} {\bibinfo {author} {\bibfnamefont {David}\ \bibnamefont
  {Poland}}, \bibinfo {author} {\bibfnamefont {Slava}\ \bibnamefont {Rychkov}},
  \ and\ \bibinfo {author} {\bibfnamefont {Alessandro}\ \bibnamefont {Vichi}},\
  }\bibfield  {title} {\enquote {\bibinfo {title} {The conformal bootstrap:
  Theory, numerical techniques, and applications},}\ }\href {\doibase
  10.1103/RevModPhys.91.015002} {\bibfield  {journal} {\bibinfo  {journal}
  {Rev. Mod. Phys.}\ }\textbf {\bibinfo {volume} {91}},\ \bibinfo {pages}
  {015002} (\bibinfo {year} {2019})}\BibitemShut {NoStop}%
\bibitem [{\citenamefont {Gaiotto}\ \emph {et~al.}(2014)\citenamefont
  {Gaiotto}, \citenamefont {Mazac},\ and\ \citenamefont
  {Paulos}}]{Gaiotto2014}%
  \BibitemOpen
  \bibfield  {author} {\bibinfo {author} {\bibfnamefont {D.}~\bibnamefont
  {Gaiotto}}, \bibinfo {author} {\bibfnamefont {D.}~\bibnamefont {Mazac}}, \
  and\ \bibinfo {author} {\bibfnamefont {M.F.}\ \bibnamefont {Paulos}},\
  }\bibfield  {title} {\enquote {\bibinfo {title} {{Bootstrapping the 3d Ising
  twist defect.}}}\ }\href {\doibase https://doi.org/10.1007/JHEP03(2014)100}
  {\bibfield  {journal} {\bibinfo  {journal} {Journal of High Energy Physics}\
  }\textbf {\bibinfo {volume} {2014}},\ \bibinfo {eid} {100} (\bibinfo {year}
  {2014})}\BibitemShut {NoStop}%
\bibitem [{\citenamefont {Gliozzi}\ \emph {et~al.}(2015)\citenamefont
  {Gliozzi}, \citenamefont {Liendo}, \citenamefont {Meineri},\ and\
  \citenamefont {Rago}}]{Gliozzi2015Boundary}%
  \BibitemOpen
  \bibfield  {author} {\bibinfo {author} {\bibfnamefont {Ferdinando}\
  \bibnamefont {Gliozzi}}, \bibinfo {author} {\bibfnamefont {Pedro}\
  \bibnamefont {Liendo}}, \bibinfo {author} {\bibfnamefont {Marco}\
  \bibnamefont {Meineri}}, \ and\ \bibinfo {author} {\bibfnamefont {Antonio}\
  \bibnamefont {Rago}},\ }\bibfield  {title} {\enquote {\bibinfo {title}
  {{Boundary and Interface CFTs from the Conformal Bootstrap}},}\ }\href
  {\doibase https://doi.org/10.1007/JHEP05(2015)036} {\bibfield  {journal}
  {\bibinfo  {journal} {Journal of High Energy Physics}\ }\textbf {\bibinfo
  {volume} {2015}},\ \bibinfo {pages} {36} (\bibinfo {year}
  {2015})}\BibitemShut {NoStop}%
\bibitem [{\citenamefont {{Padayasi}}\ \emph {et~al.}(2022)\citenamefont
  {{Padayasi}}, \citenamefont {{Krishnan}}, \citenamefont {{Metlitski}},
  \citenamefont {{Gruzberg}},\ and\ \citenamefont
  {{Meineri}}}]{Padayasi2022extraordinary}%
  \BibitemOpen
  \bibfield  {author} {\bibinfo {author} {\bibfnamefont {Jaychandran}\
  \bibnamefont {{Padayasi}}}, \bibinfo {author} {\bibfnamefont {Abijith}\
  \bibnamefont {{Krishnan}}}, \bibinfo {author} {\bibfnamefont {Max}\
  \bibnamefont {{Metlitski}}}, \bibinfo {author} {\bibfnamefont {Ilya}\
  \bibnamefont {{Gruzberg}}}, \ and\ \bibinfo {author} {\bibfnamefont {Marco}\
  \bibnamefont {{Meineri}}},\ }\bibfield  {title} {\enquote {\bibinfo {title}
  {{The extraordinary boundary transition in the 3d O(N) model via conformal
  bootstrap}},}\ }\href {\doibase 10.21468/SciPostPhys.12.6.190} {\bibfield
  {journal} {\bibinfo  {journal} {SciPost Physics}\ }\textbf {\bibinfo {volume}
  {12}},\ \bibinfo {eid} {190} (\bibinfo {year} {2022})},\ \Eprint
  {http://arxiv.org/abs/2111.03071} {arXiv:2111.03071 [cond-mat.stat-mech]}
  \BibitemShut {NoStop}%
\bibitem [{\citenamefont {{Gimenez-Grau}}\ \emph {et~al.}(2022)\citenamefont
  {{Gimenez-Grau}}, \citenamefont {{Lauria}}, \citenamefont {{Liendo}},\ and\
  \citenamefont {{van Vliet}}}]{Gimenez-Grau2022Bootstrapping}%
  \BibitemOpen
  \bibfield  {author} {\bibinfo {author} {\bibfnamefont {Aleix}\ \bibnamefont
  {{Gimenez-Grau}}}, \bibinfo {author} {\bibfnamefont {Edoardo}\ \bibnamefont
  {{Lauria}}}, \bibinfo {author} {\bibfnamefont {Pedro}\ \bibnamefont
  {{Liendo}}}, \ and\ \bibinfo {author} {\bibfnamefont {Philine}\ \bibnamefont
  {{van Vliet}}},\ }\bibfield  {title} {\enquote {\bibinfo {title}
  {{Bootstrapping line defects with O(2) global symmetry}},}\ }\href {\doibase
  10.1007/JHEP11(2022)018} {\bibfield  {journal} {\bibinfo  {journal} {Journal
  of High Energy Physics}\ }\textbf {\bibinfo {volume} {2022}},\ \bibinfo {eid}
  {18} (\bibinfo {year} {2022})},\ \Eprint {http://arxiv.org/abs/2208.11715}
  {arXiv:2208.11715 [hep-th]} \BibitemShut {NoStop}%
\bibitem [{\citenamefont {Affleck}\ and\ \citenamefont
  {Ludwig}(1991)}]{Affleck1991Universal}%
  \BibitemOpen
  \bibfield  {author} {\bibinfo {author} {\bibfnamefont {Ian}\ \bibnamefont
  {Affleck}}\ and\ \bibinfo {author} {\bibfnamefont {Andreas W.~W.}\
  \bibnamefont {Ludwig}},\ }\bibfield  {title} {\enquote {\bibinfo {title}
  {Universal noninteger ``ground-state degeneracy'' in critical quantum
  systems},}\ }\href {\doibase 10.1103/PhysRevLett.67.161} {\bibfield
  {journal} {\bibinfo  {journal} {Phys. Rev. Lett.}\ }\textbf {\bibinfo
  {volume} {67}},\ \bibinfo {pages} {161--164} (\bibinfo {year}
  {1991})}\BibitemShut {NoStop}%
\bibitem [{\citenamefont {Vojta}\ \emph {et~al.}(2000)\citenamefont {Vojta},
  \citenamefont {Buragohain},\ and\ \citenamefont
  {Sachdev}}]{Vojta2000Impurity}%
  \BibitemOpen
  \bibfield  {author} {\bibinfo {author} {\bibfnamefont {Matthias}\
  \bibnamefont {Vojta}}, \bibinfo {author} {\bibfnamefont {Chiranjeeb}\
  \bibnamefont {Buragohain}}, \ and\ \bibinfo {author} {\bibfnamefont {Subir}\
  \bibnamefont {Sachdev}},\ }\bibfield  {title} {\enquote {\bibinfo {title}
  {Quantum impurity dynamics in two-dimensional antiferromagnets and
  superconductors},}\ }\href {\doibase 10.1103/PhysRevB.61.15152} {\bibfield
  {journal} {\bibinfo  {journal} {Phys. Rev. B}\ }\textbf {\bibinfo {volume}
  {61}},\ \bibinfo {pages} {15152--15184} (\bibinfo {year} {2000})}\BibitemShut
  {NoStop}%
\bibitem [{\citenamefont {{Metlitski}}(2020)}]{Metlitski2020Boundary}%
  \BibitemOpen
  \bibfield  {author} {\bibinfo {author} {\bibfnamefont {Max~A.}\ \bibnamefont
  {{Metlitski}}},\ }\bibfield  {title} {\enquote {\bibinfo {title} {{Boundary
  criticality of the O(N) model in d = 3 critically revisited}},}\ }\href
  {\doibase 10.48550/arXiv.2009.05119} {\bibfield  {journal} {\bibinfo
  {journal} {arXiv e-prints}\ ,\ \bibinfo {eid} {arXiv:2009.05119}} (\bibinfo
  {year} {2020})},\ \Eprint {http://arxiv.org/abs/2009.05119} {arXiv:2009.05119
  [cond-mat.str-el]} \BibitemShut {NoStop}%
\bibitem [{\citenamefont {{Liu}}\ \emph {et~al.}(2021)\citenamefont {{Liu}},
  \citenamefont {{Shapourian}}, \citenamefont {{Vishwanath}},\ and\
  \citenamefont {{Metlitski}}}]{Liu2021Magnetic}%
  \BibitemOpen
  \bibfield  {author} {\bibinfo {author} {\bibfnamefont {Shang}\ \bibnamefont
  {{Liu}}}, \bibinfo {author} {\bibfnamefont {Hassan}\ \bibnamefont
  {{Shapourian}}}, \bibinfo {author} {\bibfnamefont {Ashvin}\ \bibnamefont
  {{Vishwanath}}}, \ and\ \bibinfo {author} {\bibfnamefont {Max~A.}\
  \bibnamefont {{Metlitski}}},\ }\bibfield  {title} {\enquote {\bibinfo {title}
  {{Magnetic impurities at quantum critical points: Large-N expansion and
  connections to symmetry-protected topological states}},}\ }\href {\doibase
  10.1103/PhysRevB.104.104201} {\bibfield  {journal} {\bibinfo  {journal}
  {\prb}\ }\textbf {\bibinfo {volume} {104}},\ \bibinfo {eid} {104201}
  (\bibinfo {year} {2021})},\ \Eprint {http://arxiv.org/abs/2104.15026}
  {arXiv:2104.15026 [cond-mat.str-el]} \BibitemShut {NoStop}%
\bibitem [{\citenamefont {{Krishnan}}\ and\ \citenamefont
  {{Metlitski}}(2023)}]{Krishnan2023plane}%
  \BibitemOpen
  \bibfield  {author} {\bibinfo {author} {\bibfnamefont {Abijith}\ \bibnamefont
  {{Krishnan}}}\ and\ \bibinfo {author} {\bibfnamefont {Max~A.}\ \bibnamefont
  {{Metlitski}}},\ }\bibfield  {title} {\enquote {\bibinfo {title} {{A plane
  defect in the 3d O$(N)$ model}},}\ }\href {\doibase
  10.48550/arXiv.2301.05728} {\bibfield  {journal} {\bibinfo  {journal} {arXiv
  e-prints}\ ,\ \bibinfo {eid} {arXiv:2301.05728}} (\bibinfo {year} {2023})},\
  \Eprint {http://arxiv.org/abs/2301.05728} {arXiv:2301.05728
  [cond-mat.str-el]} \BibitemShut {NoStop}%
\bibitem [{\citenamefont {{Aharony}}\ \emph {et~al.}(2023)\citenamefont
  {{Aharony}}, \citenamefont {{Cuomo}}, \citenamefont {{Komargodski}},
  \citenamefont {{Mezei}},\ and\ \citenamefont
  {{Raviv-Moshe}}}]{Aharony2023Phases}%
  \BibitemOpen
  \bibfield  {author} {\bibinfo {author} {\bibfnamefont {Ofer}\ \bibnamefont
  {{Aharony}}}, \bibinfo {author} {\bibfnamefont {Gabriel}\ \bibnamefont
  {{Cuomo}}}, \bibinfo {author} {\bibfnamefont {Zohar}\ \bibnamefont
  {{Komargodski}}}, \bibinfo {author} {\bibfnamefont {M{\'a}rk}\ \bibnamefont
  {{Mezei}}}, \ and\ \bibinfo {author} {\bibfnamefont {Avia}\ \bibnamefont
  {{Raviv-Moshe}}},\ }\bibfield  {title} {\enquote {\bibinfo {title} {{Phases
  of Wilson Lines in Conformal Field Theories}},}\ }\href {\doibase
  10.1103/PhysRevLett.130.151601} {\bibfield  {journal} {\bibinfo  {journal}
  {\prl}\ }\textbf {\bibinfo {volume} {130}},\ \bibinfo {eid} {151601}
  (\bibinfo {year} {2023})},\ \Eprint {http://arxiv.org/abs/2211.11775}
  {arXiv:2211.11775 [hep-th]} \BibitemShut {NoStop}%
\bibitem [{\citenamefont {Hanke}(2000)}]{Hanke2000}%
  \BibitemOpen
  \bibfield  {author} {\bibinfo {author} {\bibfnamefont {Andreas}\ \bibnamefont
  {Hanke}},\ }\bibfield  {title} {\enquote {\bibinfo {title} {Critical
  adsorption on defects in ising magnets and binary alloys},}\ }\href {\doibase
  10.1103/PhysRevLett.84.2180} {\bibfield  {journal} {\bibinfo  {journal}
  {Phys. Rev. Lett.}\ }\textbf {\bibinfo {volume} {84}},\ \bibinfo {pages}
  {2180--2183} (\bibinfo {year} {2000})}\BibitemShut {NoStop}%
\bibitem [{\citenamefont {Allais}\ and\ \citenamefont
  {Sachdev}(2014)}]{Andrea2014}%
  \BibitemOpen
  \bibfield  {author} {\bibinfo {author} {\bibfnamefont {Andrea}\ \bibnamefont
  {Allais}}\ and\ \bibinfo {author} {\bibfnamefont {Subir}\ \bibnamefont
  {Sachdev}},\ }\bibfield  {title} {\enquote {\bibinfo {title} {Spectral
  function of a localized fermion coupled to the wilson-fisher conformal field
  theory},}\ }\href {\doibase 10.1103/PhysRevB.90.035131} {\bibfield  {journal}
  {\bibinfo  {journal} {Phys. Rev. B}\ }\textbf {\bibinfo {volume} {90}},\
  \bibinfo {pages} {035131} (\bibinfo {year} {2014})}\BibitemShut {NoStop}%
\bibitem [{\citenamefont {{Cuomo}}\ \emph
  {et~al.}(2022{\natexlab{a}})\citenamefont {{Cuomo}}, \citenamefont
  {{Komargodski}},\ and\ \citenamefont {{Mezei}}}]{Cuomo2022magneticDefect}%
  \BibitemOpen
  \bibfield  {author} {\bibinfo {author} {\bibfnamefont {Gabriel}\ \bibnamefont
  {{Cuomo}}}, \bibinfo {author} {\bibfnamefont {Zohar}\ \bibnamefont
  {{Komargodski}}}, \ and\ \bibinfo {author} {\bibfnamefont {M{\'a}rk}\
  \bibnamefont {{Mezei}}},\ }\bibfield  {title} {\enquote {\bibinfo {title}
  {{Localized magnetic field in the O(N) model}},}\ }\href {\doibase
  10.1007/JHEP02(2022)134} {\bibfield  {journal} {\bibinfo  {journal} {Journal
  of High Energy Physics}\ }\textbf {\bibinfo {volume} {2022}},\ \bibinfo {eid}
  {134} (\bibinfo {year} {2022}{\natexlab{a}})},\ \Eprint
  {http://arxiv.org/abs/2112.10634} {arXiv:2112.10634 [hep-th]} \BibitemShut
  {NoStop}%
\bibitem [{\citenamefont {{Allais}}(2014)}]{Allais2014}%
  \BibitemOpen
  \bibfield  {author} {\bibinfo {author} {\bibfnamefont {Andrea}\ \bibnamefont
  {{Allais}}},\ }\bibfield  {title} {\enquote {\bibinfo {title} {{Magnetic
  defect line in a critical Ising bath}},}\ }\href {\doibase
  10.48550/arXiv.1412.3449} {\bibfield  {journal} {\bibinfo  {journal} {arXiv
  e-prints}\ ,\ \bibinfo {eid} {arXiv:1412.3449}} (\bibinfo {year} {2014})},\
  \Eprint {http://arxiv.org/abs/1412.3449} {arXiv:1412.3449 [cond-mat.str-el]}
  \BibitemShut {NoStop}%
\bibitem [{\citenamefont {Parisen~Toldin}\ \emph {et~al.}(2017)\citenamefont
  {Parisen~Toldin}, \citenamefont {Assaad},\ and\ \citenamefont
  {Wessel}}]{Assaad2017}%
  \BibitemOpen
  \bibfield  {author} {\bibinfo {author} {\bibfnamefont {Francesco}\
  \bibnamefont {Parisen~Toldin}}, \bibinfo {author} {\bibfnamefont {Fakher~F.}\
  \bibnamefont {Assaad}}, \ and\ \bibinfo {author} {\bibfnamefont {Stefan}\
  \bibnamefont {Wessel}},\ }\bibfield  {title} {\enquote {\bibinfo {title}
  {Critical behavior in the presence of an order-parameter pinning field},}\
  }\href {\doibase 10.1103/PhysRevB.95.014401} {\bibfield  {journal} {\bibinfo
  {journal} {Phys. Rev. B}\ }\textbf {\bibinfo {volume} {95}},\ \bibinfo
  {pages} {014401} (\bibinfo {year} {2017})}\BibitemShut {NoStop}%
\bibitem [{\citenamefont {{Parisen Toldin}}\ and\ \citenamefont
  {{Metlitski}}(2022)}]{Toldin2022Boundary}%
  \BibitemOpen
  \bibfield  {author} {\bibinfo {author} {\bibfnamefont {Francesco}\
  \bibnamefont {{Parisen Toldin}}}\ and\ \bibinfo {author} {\bibfnamefont
  {Max~A.}\ \bibnamefont {{Metlitski}}},\ }\bibfield  {title} {\enquote
  {\bibinfo {title} {{Boundary Criticality of the 3D O(N ) Model: From Normal
  to Extraordinary}},}\ }\href {\doibase 10.1103/PhysRevLett.128.215701}
  {\bibfield  {journal} {\bibinfo  {journal} {\prl}\ }\textbf {\bibinfo
  {volume} {128}},\ \bibinfo {eid} {215701} (\bibinfo {year} {2022})},\ \Eprint
  {http://arxiv.org/abs/2111.03613} {arXiv:2111.03613 [cond-mat.stat-mech]}
  \BibitemShut {NoStop}%
\bibitem [{\citenamefont {{Cuomo}}\ \emph
  {et~al.}(2022{\natexlab{b}})\citenamefont {{Cuomo}}, \citenamefont
  {{Komargodski}},\ and\ \citenamefont {{Raviv-Moshe}}}]{Cuomo2022gtheorem}%
  \BibitemOpen
  \bibfield  {author} {\bibinfo {author} {\bibfnamefont {Gabriel}\ \bibnamefont
  {{Cuomo}}}, \bibinfo {author} {\bibfnamefont {Zohar}\ \bibnamefont
  {{Komargodski}}}, \ and\ \bibinfo {author} {\bibfnamefont {Avia}\
  \bibnamefont {{Raviv-Moshe}}},\ }\bibfield  {title} {\enquote {\bibinfo
  {title} {{Renormalization Group Flows on Line Defects}},}\ }\href {\doibase
  10.1103/PhysRevLett.128.021603} {\bibfield  {journal} {\bibinfo  {journal}
  {\prl}\ }\textbf {\bibinfo {volume} {128}},\ \bibinfo {eid} {021603}
  (\bibinfo {year} {2022}{\natexlab{b}})},\ \Eprint
  {http://arxiv.org/abs/2108.01117} {arXiv:2108.01117 [hep-th]} \BibitemShut
  {NoStop}%
\bibitem [{\citenamefont {{Casini}}\ \emph {et~al.}(2023)\citenamefont
  {{Casini}}, \citenamefont {{Landea}},\ and\ \citenamefont
  {{Torroba}}}]{Casini2023Entropic}%
  \BibitemOpen
  \bibfield  {author} {\bibinfo {author} {\bibfnamefont {Horacio}\ \bibnamefont
  {{Casini}}}, \bibinfo {author} {\bibfnamefont {Ignacio~Salazar}\ \bibnamefont
  {{Landea}}}, \ and\ \bibinfo {author} {\bibfnamefont {Gonzalo}\ \bibnamefont
  {{Torroba}}},\ }\bibfield  {title} {\enquote {\bibinfo {title} {{Entropic g
  Theorem in General Spacetime Dimensions}},}\ }\href {\doibase
  10.1103/PhysRevLett.130.111603} {\bibfield  {journal} {\bibinfo  {journal}
  {\prl}\ }\textbf {\bibinfo {volume} {130}},\ \bibinfo {eid} {111603}
  (\bibinfo {year} {2023})},\ \Eprint {http://arxiv.org/abs/2212.10575}
  {arXiv:2212.10575 [hep-th]} \BibitemShut {NoStop}%
\bibitem [{\citenamefont {{Friedan}}\ and\ \citenamefont
  {{Konechny}}(2004)}]{Friedan2004Boundary}%
  \BibitemOpen
  \bibfield  {author} {\bibinfo {author} {\bibfnamefont {Daniel}\ \bibnamefont
  {{Friedan}}}\ and\ \bibinfo {author} {\bibfnamefont {Anatoly}\ \bibnamefont
  {{Konechny}}},\ }\bibfield  {title} {\enquote {\bibinfo {title} {{Boundary
  Entropy of One-Dimensional Quantum Systems at Low Temperature}},}\ }\href
  {\doibase 10.1103/PhysRevLett.93.030402} {\bibfield  {journal} {\bibinfo
  {journal} {\prl}\ }\textbf {\bibinfo {volume} {93}},\ \bibinfo {eid} {030402}
  (\bibinfo {year} {2004})},\ \Eprint {http://arxiv.org/abs/hep-th/0312197}
  {arXiv:hep-th/0312197 [hep-th]} \BibitemShut {NoStop}%
\bibitem [{\citenamefont {{Casini}}\ \emph {et~al.}(2016)\citenamefont
  {{Casini}}, \citenamefont {{Landea}},\ and\ \citenamefont
  {{Torroba}}}]{Casini20162Dg}%
  \BibitemOpen
  \bibfield  {author} {\bibinfo {author} {\bibfnamefont {Horacio}\ \bibnamefont
  {{Casini}}}, \bibinfo {author} {\bibfnamefont {Ignacio~Salazar}\ \bibnamefont
  {{Landea}}}, \ and\ \bibinfo {author} {\bibfnamefont {Gonzalo}\ \bibnamefont
  {{Torroba}}},\ }\bibfield  {title} {\enquote {\bibinfo {title} {{The
  g-theorem and quantum information theory}},}\ }\href {\doibase
  10.1007/JHEP10(2016)140} {\bibfield  {journal} {\bibinfo  {journal} {Journal
  of High Energy Physics}\ }\textbf {\bibinfo {volume} {2016}},\ \bibinfo {eid}
  {140} (\bibinfo {year} {2016})},\ \Eprint {http://arxiv.org/abs/1607.00390}
  {arXiv:1607.00390 [hep-th]} \BibitemShut {NoStop}%
\bibitem [{\citenamefont {Zhu}\ \emph {et~al.}(2023)\citenamefont {Zhu},
  \citenamefont {Han}, \citenamefont {Huffman}, \citenamefont {Hofmann},\ and\
  \citenamefont {He}}]{ZHHHH2022}%
  \BibitemOpen
  \bibfield  {author} {\bibinfo {author} {\bibfnamefont {W.}~\bibnamefont
  {Zhu}}, \bibinfo {author} {\bibfnamefont {Chao}\ \bibnamefont {Han}},
  \bibinfo {author} {\bibfnamefont {Emilie}\ \bibnamefont {Huffman}}, \bibinfo
  {author} {\bibfnamefont {Johannes~S.}\ \bibnamefont {Hofmann}}, \ and\
  \bibinfo {author} {\bibfnamefont {Yin-Chen}\ \bibnamefont {He}},\ }\bibfield
  {title} {\enquote {\bibinfo {title} {Uncovering conformal symmetry in the 3d
  ising transition: State-operator correspondence from a quantum fuzzy sphere
  regularization},}\ }\href {\doibase 10.1103/PhysRevX.13.021009} {\bibfield
  {journal} {\bibinfo  {journal} {Phys. Rev. X}\ }\textbf {\bibinfo {volume}
  {13}},\ \bibinfo {pages} {021009} (\bibinfo {year} {2023})}\BibitemShut
  {NoStop}%
\bibitem [{\citenamefont {Hu}\ \emph {et~al.}(2023)\citenamefont {Hu},
  \citenamefont {He},\ and\ \citenamefont {Zhu}}]{hu2023operator}%
  \BibitemOpen
  \bibfield  {author} {\bibinfo {author} {\bibfnamefont {Liangdong}\
  \bibnamefont {Hu}}, \bibinfo {author} {\bibfnamefont {Yin-Chen}\ \bibnamefont
  {He}}, \ and\ \bibinfo {author} {\bibfnamefont {W.}~\bibnamefont {Zhu}},\
  }\bibfield  {title} {\enquote {\bibinfo {title} {Operator product expansion
  coefficients of the 3d ising criticality via quantum fuzzy spheres},}\ }\href
  {\doibase 10.1103/PhysRevLett.131.031601} {\bibfield  {journal} {\bibinfo
  {journal} {Phys. Rev. Lett.}\ }\textbf {\bibinfo {volume} {131}},\ \bibinfo
  {pages} {031601} (\bibinfo {year} {2023})}\BibitemShut {NoStop}%
\bibitem [{\citenamefont {Han}\ \emph {et~al.}(2023)\citenamefont {Han},
  \citenamefont {Hu}, \citenamefont {Zhu},\ and\ \citenamefont
  {He}}]{Han2023Conformal}%
  \BibitemOpen
  \bibfield  {author} {\bibinfo {author} {\bibfnamefont {Chao}\ \bibnamefont
  {Han}}, \bibinfo {author} {\bibfnamefont {Liangdong}\ \bibnamefont {Hu}},
  \bibinfo {author} {\bibfnamefont {W.}~\bibnamefont {Zhu}}, \ and\ \bibinfo
  {author} {\bibfnamefont {Yin-Chen}\ \bibnamefont {He}},\ }\bibfield  {title}
  {\enquote {\bibinfo {title} {{Conformal four-point correlators of the 3D
  Ising transition via the quantum fuzzy sphere}},}\ }\href@noop {} {\
  (\bibinfo {year} {2023})},\ \Eprint {http://arxiv.org/abs/2306.04681}
  {arXiv:2306.04681 [cond-mat.stat-mech]} \BibitemShut {NoStop}%
\bibitem [{\citenamefont {{Zhou}}\ \emph {et~al.}(2023)\citenamefont {{Zhou}},
  \citenamefont {{Hu}}, \citenamefont {{Zhu}},\ and\ \citenamefont
  {{He}}}]{Zhou2023SO5}%
  \BibitemOpen
  \bibfield  {author} {\bibinfo {author} {\bibfnamefont {Zheng}\ \bibnamefont
  {{Zhou}}}, \bibinfo {author} {\bibfnamefont {Liangdong}\ \bibnamefont
  {{Hu}}}, \bibinfo {author} {\bibfnamefont {W.}~\bibnamefont {{Zhu}}}, \ and\
  \bibinfo {author} {\bibfnamefont {Yin-Chen}\ \bibnamefont {{He}}},\
  }\bibfield  {title} {\enquote {\bibinfo {title} {{The $\mathrm{SO}(5)$
  Deconfined Phase Transition under the Fuzzy Sphere Microscope: Approximate
  Conformal Symmetry, Pseudo-Criticality, and Operator Spectrum}},}\ }\href
  {\doibase 10.48550/arXiv.2306.16435} {\bibfield  {journal} {\bibinfo
  {journal} {arXiv e-prints}\ ,\ \bibinfo {eid} {arXiv:2306.16435}} (\bibinfo
  {year} {2023})},\ \Eprint {http://arxiv.org/abs/2306.16435} {arXiv:2306.16435
  [cond-mat.str-el]} \BibitemShut {NoStop}%
\bibitem [{\citenamefont {{Nishioka}}\ \emph {et~al.}(2023)\citenamefont
  {{Nishioka}}, \citenamefont {{Okuyama}},\ and\ \citenamefont
  {{Shimamori}}}]{Nishioka2023}%
  \BibitemOpen
  \bibfield  {author} {\bibinfo {author} {\bibfnamefont {Tatsuma}\ \bibnamefont
  {{Nishioka}}}, \bibinfo {author} {\bibfnamefont {Yoshitaka}\ \bibnamefont
  {{Okuyama}}}, \ and\ \bibinfo {author} {\bibfnamefont {Soichiro}\
  \bibnamefont {{Shimamori}}},\ }\bibfield  {title} {\enquote {\bibinfo {title}
  {{The epsilon expansion of the O(N) model with line defect from conformal
  field theory}},}\ }\href {\doibase 10.1007/JHEP03(2023)203} {\bibfield
  {journal} {\bibinfo  {journal} {Journal of High Energy Physics}\ }\textbf
  {\bibinfo {volume} {2023}},\ \bibinfo {eid} {203} (\bibinfo {year} {2023})},\
  \Eprint {http://arxiv.org/abs/2212.04076} {arXiv:2212.04076 [hep-th]}
  \BibitemShut {NoStop}%
\bibitem [{\citenamefont {{Bianchi}}\ \emph {et~al.}(2023)\citenamefont
  {{Bianchi}}, \citenamefont {{Bonomi}},\ and\ \citenamefont {{de
  Sabbata}}}]{Bianchi2023Analytic}%
  \BibitemOpen
  \bibfield  {author} {\bibinfo {author} {\bibfnamefont {Lorenzo}\ \bibnamefont
  {{Bianchi}}}, \bibinfo {author} {\bibfnamefont {Davide}\ \bibnamefont
  {{Bonomi}}}, \ and\ \bibinfo {author} {\bibfnamefont {Elia}\ \bibnamefont
  {{de Sabbata}}},\ }\bibfield  {title} {\enquote {\bibinfo {title} {{Analytic
  bootstrap for the localized magnetic field}},}\ }\href {\doibase
  10.1007/JHEP04(2023)069} {\bibfield  {journal} {\bibinfo  {journal} {Journal
  of High Energy Physics}\ }\textbf {\bibinfo {volume} {2023}},\ \bibinfo {eid}
  {69} (\bibinfo {year} {2023})},\ \Eprint {http://arxiv.org/abs/2212.02524}
  {arXiv:2212.02524 [hep-th]} \BibitemShut {NoStop}%
\bibitem [{\citenamefont {{Gimenez-Grau}}(2022)}]{Gimenez-Grau2022Probing}%
  \BibitemOpen
  \bibfield  {author} {\bibinfo {author} {\bibfnamefont {Aleix}\ \bibnamefont
  {{Gimenez-Grau}}},\ }\bibfield  {title} {\enquote {\bibinfo {title} {{Probing
  magnetic line defects with two-point functions}},}\ }\href {\doibase
  10.48550/arXiv.2212.02520} {\bibfield  {journal} {\bibinfo  {journal} {arXiv
  e-prints}\ ,\ \bibinfo {eid} {arXiv:2212.02520}} (\bibinfo {year} {2022})},\
  \Eprint {http://arxiv.org/abs/2212.02520} {arXiv:2212.02520 [hep-th]}
  \BibitemShut {NoStop}%
\bibitem [{\citenamefont {{Pannell}}\ and\ \citenamefont
  {{Stergiou}}(2023)}]{Pannell2023Line}%
  \BibitemOpen
  \bibfield  {author} {\bibinfo {author} {\bibfnamefont {William~H.}\
  \bibnamefont {{Pannell}}}\ and\ \bibinfo {author} {\bibfnamefont {Andreas}\
  \bibnamefont {{Stergiou}}},\ }\bibfield  {title} {\enquote {\bibinfo {title}
  {{Line defect RG flows in the {\ensuremath{\varepsilon}} expansion}},}\
  }\href {\doibase 10.1007/JHEP06(2023)186} {\bibfield  {journal} {\bibinfo
  {journal} {Journal of High Energy Physics}\ }\textbf {\bibinfo {volume}
  {2023}},\ \bibinfo {eid} {186} (\bibinfo {year} {2023})},\ \Eprint
  {http://arxiv.org/abs/2302.14069} {arXiv:2302.14069 [hep-th]} \BibitemShut
  {NoStop}%
\bibitem [{SM()}]{SM}%
  \BibitemOpen
  \href@noop {} {\bibinfo  {journal} {Supplementary material}\ }\BibitemShut
  {NoStop}%
\bibitem [{\citenamefont {Cardy}(1984{\natexlab{b}})}]{Cardy1984}%
  \BibitemOpen
\bibfield  {journal} {  }\bibfield  {author} {\bibinfo {author} {\bibfnamefont
  {J~L}\ \bibnamefont {Cardy}},\ }\bibfield  {title} {\enquote {\bibinfo
  {title} {Conformal invariance and universality in finite-size scaling},}\
  }\href {\doibase 10.1088/0305-4470/17/7/003} {\bibfield  {journal} {\bibinfo
  {journal} {Journal of Physics A: Mathematical and General}\ }\textbf
  {\bibinfo {volume} {17}},\ \bibinfo {pages} {L385--L387} (\bibinfo {year}
  {1984}{\natexlab{b}})}\BibitemShut {NoStop}%
\bibitem [{\citenamefont {Cardy}(1985)}]{Cardy1985}%
  \BibitemOpen
  \bibfield  {author} {\bibinfo {author} {\bibfnamefont {J~L}\ \bibnamefont
  {Cardy}},\ }\bibfield  {title} {\enquote {\bibinfo {title} {Universal
  amplitudes in finite-size scaling: generalisation to arbitrary
  dimensionality},}\ }\href {\doibase 10.1088/0305-4470/18/13/005} {\bibfield
  {journal} {\bibinfo  {journal} {Journal of Physics A: Mathematical and
  General}\ }\textbf {\bibinfo {volume} {18}},\ \bibinfo {pages} {L757--L760}
  (\bibinfo {year} {1985})}\BibitemShut {NoStop}%
\bibitem [{\citenamefont {Wu}\ and\ \citenamefont
  {Yang}(1976)}]{WuYangmonopole}%
  \BibitemOpen
  \bibfield  {author} {\bibinfo {author} {\bibfnamefont {Tai~Tsun}\
  \bibnamefont {Wu}}\ and\ \bibinfo {author} {\bibfnamefont {Chen~Ning}\
  \bibnamefont {Yang}},\ }\bibfield  {title} {\enquote {\bibinfo {title} {Dirac
  monopole without strings: monopole harmonics},}\ }\href@noop {} {\bibfield
  {journal} {\bibinfo  {journal} {Nuclear Physics B}\ }\textbf {\bibinfo
  {volume} {107}},\ \bibinfo {pages} {365--380} (\bibinfo {year}
  {1976})}\BibitemShut {NoStop}%
\bibitem [{\citenamefont {Madore}(1992)}]{madore1992fuzzy}%
  \BibitemOpen
  \bibfield  {author} {\bibinfo {author} {\bibfnamefont {John}\ \bibnamefont
  {Madore}},\ }\bibfield  {title} {\enquote {\bibinfo {title} {The fuzzy
  sphere},}\ }\href@noop {} {\bibfield  {journal} {\bibinfo  {journal}
  {Classical and Quantum Gravity}\ }\textbf {\bibinfo {volume} {9}},\ \bibinfo
  {pages} {69} (\bibinfo {year} {1992})}\BibitemShut {NoStop}%
\bibitem [{\citenamefont {Law}(2001)}]{LAW2001159}%
  \BibitemOpen
  \bibfield  {author} {\bibinfo {author} {\bibfnamefont {Bruce~M.}\
  \bibnamefont {Law}},\ }\bibfield  {title} {\enquote {\bibinfo {title}
  {Wetting, adsorption and surface critical phenomena},}\ }\href {\doibase
  https://doi.org/10.1016/S0079-6816(00)00025-3} {\bibfield  {journal}
  {\bibinfo  {journal} {Progress in Surface Science}\ }\textbf {\bibinfo
  {volume} {66}},\ \bibinfo {pages} {159--216} (\bibinfo {year}
  {2001})}\BibitemShut {NoStop}%
\bibitem [{\citenamefont {Fisher}\ and\ \citenamefont
  {de~Gennes}()}]{Fisher_Gennes}%
  \BibitemOpen
  \bibfield  {author} {\bibinfo {author} {\bibfnamefont {Michael~E.}\
  \bibnamefont {Fisher}}\ and\ \bibinfo {author} {\bibfnamefont
  {Pierre-Gilles}\ \bibnamefont {de~Gennes}},\ }\enquote {\bibinfo {title}
  {Phénomènes aux parois dans un mélange binaire critique},}\ in\ \href
  {\doibase 10.1142/9789812564849_0025} {\emph {\bibinfo {booktitle} {Simple
  Views on Condensed Matter}}},\ pp.\ \bibinfo {pages} {237--241}\BibitemShut
  {NoStop}%
\bibitem [{\citenamefont {{Lao}}\ and\ \citenamefont
  {{Rychkov}}(2023)}]{Rychkov2023Icosahedron}%
  \BibitemOpen
  \bibfield  {author} {\bibinfo {author} {\bibfnamefont {Bing-Xin}\
  \bibnamefont {{Lao}}}\ and\ \bibinfo {author} {\bibfnamefont {Slava}\
  \bibnamefont {{Rychkov}}},\ }\bibfield  {title} {\enquote {\bibinfo {title}
  {{3D Ising CFT and Exact Diagonalization on Icosahedron}},}\ }\href {\doibase
  10.48550/arXiv.2307.02540} {\bibfield  {journal} {\bibinfo  {journal} {arXiv
  e-prints}\ ,\ \bibinfo {eid} {arXiv:2307.02540}} (\bibinfo {year} {2023})},\
  \Eprint {http://arxiv.org/abs/2307.02540} {arXiv:2307.02540 [hep-th]}
  \BibitemShut {NoStop}%
\bibitem [{\citenamefont {Haldane}(1983)}]{Sphere_LL_Haldane}%
  \BibitemOpen
  \bibfield  {author} {\bibinfo {author} {\bibfnamefont {F.~D.~M.}\
  \bibnamefont {Haldane}},\ }\bibfield  {title} {\enquote {\bibinfo {title}
  {Fractional quantization of the hall effect: A hierarchy of incompressible
  quantum fluid states},}\ }\href {\doibase 10.1103/PhysRevLett.51.605}
  {\bibfield  {journal} {\bibinfo  {journal} {Phys. Rev. Lett.}\ }\textbf
  {\bibinfo {volume} {51}},\ \bibinfo {pages} {605--608} (\bibinfo {year}
  {1983})}\BibitemShut {NoStop}%
\bibitem [{\citenamefont {White}(1992)}]{SWhite1992}%
  \BibitemOpen
  \bibfield  {author} {\bibinfo {author} {\bibfnamefont {Steven~R.}\
  \bibnamefont {White}},\ }\bibfield  {title} {\enquote {\bibinfo {title}
  {Density matrix formulation for quantum renormalization groups},}\ }\href
  {\doibase 10.1103/PhysRevLett.69.2863} {\bibfield  {journal} {\bibinfo
  {journal} {Phys. Rev. Lett.}\ }\textbf {\bibinfo {volume} {69}},\ \bibinfo
  {pages} {2863--2866} (\bibinfo {year} {1992})}\BibitemShut {NoStop}%
\bibitem [{\citenamefont {Feiguin}\ \emph {et~al.}(2008)\citenamefont
  {Feiguin}, \citenamefont {Rezayi}, \citenamefont {Nayak},\ and\ \citenamefont
  {Das~Sarma}}]{Feiguin2008}%
  \BibitemOpen
  \bibfield  {author} {\bibinfo {author} {\bibfnamefont {A.~E.}\ \bibnamefont
  {Feiguin}}, \bibinfo {author} {\bibfnamefont {E.}~\bibnamefont {Rezayi}},
  \bibinfo {author} {\bibfnamefont {C.}~\bibnamefont {Nayak}}, \ and\ \bibinfo
  {author} {\bibfnamefont {S.}~\bibnamefont {Das~Sarma}},\ }\bibfield  {title}
  {\enquote {\bibinfo {title} {Density matrix renormalization group study of
  incompressible fractional quantum hall states},}\ }\href {\doibase
  10.1103/PhysRevLett.100.166803} {\bibfield  {journal} {\bibinfo  {journal}
  {Phys. Rev. Lett.}\ }\textbf {\bibinfo {volume} {100}},\ \bibinfo {pages}
  {166803} (\bibinfo {year} {2008})}\BibitemShut {NoStop}%
\bibitem [{\citenamefont {Fishman}\ \emph {et~al.}(2020)\citenamefont
  {Fishman}, \citenamefont {White},\ and\ \citenamefont
  {Stoudenmire}}]{fishman2020itensor}%
  \BibitemOpen
  \bibfield  {author} {\bibinfo {author} {\bibfnamefont {Matthew}\ \bibnamefont
  {Fishman}}, \bibinfo {author} {\bibfnamefont {Steven~R.}\ \bibnamefont
  {White}}, \ and\ \bibinfo {author} {\bibfnamefont {E.~Miles}\ \bibnamefont
  {Stoudenmire}},\ }\href@noop {} {\enquote {\bibinfo {title} {The
  \mbox{ITensor} software library for tensor network calculations},}\ }
  (\bibinfo {year} {2020}),\ \Eprint {http://arxiv.org/abs/2007.14822}
  {arXiv:2007.14822 [cs.MS]} \BibitemShut {NoStop}%
\bibitem [{\citenamefont {Zhou}\ \emph {et~al.}(2024)\citenamefont {Zhou},
  \citenamefont {Gaiotto}, \citenamefont {He},\ and\ \citenamefont
  {Zou}}]{zhou2024gfunction}%
  \BibitemOpen
  \bibfield  {author} {\bibinfo {author} {\bibfnamefont {Zheng}\ \bibnamefont
  {Zhou}}, \bibinfo {author} {\bibfnamefont {Davide}\ \bibnamefont {Gaiotto}},
  \bibinfo {author} {\bibfnamefont {Yin-Chen}\ \bibnamefont {He}}, \ and\
  \bibinfo {author} {\bibfnamefont {Yijian}\ \bibnamefont {Zou}},\ }\href@noop
  {} {\enquote {\bibinfo {title} {The $g$-function and defect changing
  operators from wavefunction overlap on a fuzzy sphere},}\ } (\bibinfo {year}
  {2024}),\ \Eprint {http://arxiv.org/abs/2401.00039} {arXiv:2401.00039
  [hep-th]} \BibitemShut {NoStop}%
\end{thebibliography}%



\clearpage

\onecolumngrid 
\newpage

\begin{center}
\textbf{\large Supplementary Information for ``Solving Conformal Defects in 3D Conformal Field Theory using Fuzzy Sphere Regularization''}
\end{center}
{\bf Supplementary Information}\\
Supplementary information contains:\\
 More details to support the discussion in the main text.\\
Fig. S1. Energy gap scaling at the critical point.\\
Fig. S2. Finite-size correction of scaling dimensions for the lower primaries.\\
Fig. S3. Correlators related to $\epsilon$. \\
Fig. S4. Correlators between the displacement operator and bulk operator. \\
Fig .S5. Conformal tower of defect primaries in $h_d\rightarrow\infty$. \\
TABLE S1. The scaling dimensions of primaries in sector $L_z = 0, 1$.\\
TABLE S2. The scaling dimensions of primary fields under infinite and finite defect strengths. \\
\tableofcontents

\newpage

\setcounter{subsection}{0}
\setcounter{equation}{0}
\setcounter{figure}{0}
\setcounter{table}{0}
\renewcommand{\tablename}{Supplementary Table}

\renewcommand{\figurename}{Supplementary Figure}

\appendix
In this supplementary material, we will show more details to support the discussion in the main text. In Supplementary Note 1, we discuss the defect CFT correlators on the cylinder $S^2\times \mathbb R$. In Supplementary Note 2, we present an error analysis of the scaling dimensions of defect primaries.
In Supplementary Note 3, we provide in-depth analysis of the attractive fixed point induced by the line defect.
In Supplementary Note 4, we show the operator product expansion (OPE) related to the primary $\epsilon$. 
In Supplementary Note 5 we present the computation of Zamolodchikov norm regarding to the displacement operator. 
In Supplementary Note 6, we discuss the physics of defect CFT in the limit of $h_d \rightarrow \infty$. 
	
\section{Supplementary Note 1: Bulk to defect correlators}
In this section, we present the correlators of primary operators in the dCFT on the cylinder $S^{2}\times \mathbb R$ by using the state-operator correspondence. As discussed in the main text, by introducing of a flat $p$-dimensional defect breaks the global conformal symmetry $SO(4,1)$ into $SO(p+1,1)\times SO(3-p)$.  
Making use of state-operator correspondence, the bulk-defect (scalar-scalar) correlator $\langle O_1(x) \hat O_2(0) \rangle = \frac{b_{O_1 \hat O_2}}{|x_\perp|^{\Delta_1- \hat \Delta_2} |x|^{2\hat \Delta_2}}  $ can be mapped to
    \begin{equation}
    \langle \hat{1}|O_1(x)|\hat O_2\rangle =\frac{b_{O_1 \hat O_2}}{|x_\perp|^{\Delta_1- \hat \Delta_2} |x|^{2\hat \Delta_2}} 
    \end{equation}
    where $|\hat{1} \rangle $
    is the vacuum state of the dCFT. 
    
    Next, we apply the Weyl transformation $\tau = R \ln r$ to map the coordinates $(r,\theta,\varphi)$ in $\mathbb R^3$ to $(\tau,\theta,\varphi)$ in $S^{2}\times\mathbb R$, where $R$ is the radius of $S^{2}$. Under the Weyl transformation, the operator transforms as:
    
    \begin{equation}
     \phi(r,\theta,\varphi)\rightarrow  \phi(\tau,\theta,\varphi) = \Lambda(r,\theta,\varphi)^{\Delta/2}  \phi(r, \theta,\varphi),
    \end{equation}
    where $\Lambda = R^{-2}e^\frac{2\tau}{R}$ represents the scale factor of this transformation. Substituting this into the correlator and setting $\tau=0$, we get:
    
    \begin{equation} \label{eq:bulktodef_correlator}
    \langle \hat{1}|O_1(\tau=0,\theta,\varphi)|\hat O_2\rangle = \langle \hat{1}|\Lambda(r,\theta,\varphi)^{\Delta_1/2} O_1(r, \theta,\varphi)|\hat O_2\rangle\big|_{r=1} = R^{-\Delta_1}\frac{b_{O_1\hat O_2}}{|\sin\theta|^{\Delta_1-\hat \Delta_{2}} }
    \end{equation}
    On the hand, we have $\langle {1}|O_1(\tau=0,\theta,\varphi)| O_1\rangle=R^{-\Delta_1}$ for the bulk CFT, so we finally have,
    \begin{equation}
G_{O_1\hat{O}_2}\equiv \frac{ \langle \hat 1 | O_1(\tau=0, \theta,\varphi) |\hat O_2\rangle}{\langle 1| O_1(\tau=0, \theta,\varphi) |O_{1}\rangle} = \frac{b_{O_1 \hat O_2}}{(\sin \theta)^{\Delta_1-\hat \Delta_2}}
    \end{equation}

    Lastly, it is noting that, for the flat line defect ($d=1$) breaking the symmetry into $SO(2,1)\times SO(2)$, the defect operator can have a non-trivial $SO(2)$ quantum number $L_z=m$. For such a defect operator, we have,  
    \begin{equation}
G_{O_1\hat{O}_2}\equiv \frac{  e^{-im\varphi} \langle \hat{1} | O_1(\tau=0, \theta,\varphi) |\hat O_2, L_z=m\rangle}{\langle 1| O_1(\tau=0, \theta,\varphi) |O_{1}\rangle} = \frac{b_{O_1 \hat O_2}}{(\sin \theta)^{\Delta_1-\hat \Delta_2}},
    \end{equation}
where the bulk operator $O_1$ is still a Lorentz scalar.
    
\section{Supplementary Note 2: Error analysis}
\subsection{Scaling dimension}
In this section, we provide a detailed analysis of the scaling dimensions of primaries and offer a way to estimate the error of the obtained numerical data.  
In general, in this work, we use two different methods to obtain the scaling dimensions of the defect primaries, which gives consistent results. 

These three methods are described as below.
\begin{enumerate}

    \item Since the scaling dimension of the displacement operator is expected to be $\Delta_{\hat D}=2$, we can set the dimension of the displacement operator to $\Delta_{\hat D} = 2$ and rescale the energy spectrum. The obtained results of low-lying defect primaries are shown in the first line of Supplementary Table \ref{stab:primary}.
    
    \item We assume that the presence of defect does not affect the velocity of spectra, so we let $v_{\text{def}}=v_{\text{bulk}}$. 
    So we determine the scaling dimensions $\Delta_{\hat O}$ through
    $$
         E_{\hat O}-E_{\hat{1}} = \frac{\Delta_{\hat O}}{R}v_{\text{bulk}},
    $$
    where $v_{\text{bulk}}$ is determined by the bulk Ising CFT \cite{ZHHHH2022}. That is, after extracting the velocity $v_{\text{bulk}}$, the scaling dimensions of the dCFT is
    $$
        \Delta_{\hat O} = \frac{E_{\hat O}-E_{\hat{1}}}{v_{\text{bulk}}} R = \frac{E_{\hat O}-E_{\hat{1}}}{E_\sigma-E_0}\Delta_{\sigma},
    $$ 
    where $\Delta_\sigma=0.518149$ is for bulk primary $\sigma$ field.
    The results obtained in this way are displayed in the second line of Supplementary Table \ref{stab:primary}. The consistency of methods 1 and 2 is strong evidence of $v_{\text{def}}=v_{\text{bulk}}$. Throughout this article, we employ this method to compute the scaling dimensions.
    
\end{enumerate}
The summary of the scaling dimensions obtained by the above two methods is shown in Supplementary Table \ref{stab:primary}. 

Furthermore, to estimate the scaling dimensions in the thermodynamic limit, we apply a finite-size extrapolation analysis based on method 2. 
Generally, we fit the scaling dimensions of primaries (obtained on finite-size $N$) using the following form (see section C):
\begin{equation}\label{seq:s6}
    \Delta_{\hat O}^{\text{Method 2}}(N) \approx \Delta_{\hat O}
    +\frac{b}{R^{\Delta_{\hat \phi}-1}}
    +\frac{c}{R^{\Delta_{\hat \phi'}-1}}
    +\text{higher order corrections}
\end{equation}
where $b,c$ are non-universal parameters, $R\sim \sqrt{N}$. 
The finite-size scaling for several typical primaries are shown in Supplementary Figure \ref{sfig:deltaBulkVFit}. 
The scaling dimension in the thermodynamic limit $\Delta_{\hat O}$ can be extracted in this way.

At last, the relative error is estimated through the following comparison. The numerical error of the fitting process in Supplementary Eq. (\ref{seq:s6}) is $\delta \overline{\Delta}_{\hat O}$, which is determined by comparing the different fitting processes using various finite system sizes (five large system sizes are always included). We also compare different methods to determine the relative errors, e.g. difference between $\Delta_{\hat O}$ and $\Delta_{\hat O}^{\text{Method 1}}(N=36), \Delta_{\hat O}^{\text{Method 2}}(N=36)$ as the relative error. Finally,
we use the maximum value of these estimates as the relative error:
\begin{equation}\label{seq:s7}
    \delta \Delta_{\hat O} = Max\{ \delta \overline{\Delta}_{\hat O} ,|\Delta_{\hat O}-\Delta_{\hat O}^{\text{Method 1}}(N=36)|, |\Delta_{\hat O}-\Delta_{\hat O}^{\text{Method 2}}(N=36)| \}.
\end{equation}
We think this error represents the maximal relative error of the obtained scaling dimensionsin our numerical calculations.

Finally, the numerical estimation based on the finite-size extrapolation and corresponding error bars are presented in the Table I in the main text.

\begin{table}
\setlength{\tabcolsep}{0.2cm}
\renewcommand{\arraystretch}{1.4}
    \centering
    \caption{The scaling dimensions of primaries in sector $L_z=0,1$. The first two lines correspond to method 1-2 in size $N=36$.
    } \label{stab:primary}
\begin{tabular}{c|ccc|ccc} 
\hline\hline
&\multicolumn{3}{c|}{$L_z=0$}& \multicolumn{3}{c}{$L_z=1$} \\
    \hline
 & $\hat \phi$ & $\hat \phi'$ & $\hat \phi''$  & $\hat D$ & $\hat \phi_1$ & $\hat \phi_1'$\\
$\Delta_{\hat O}^{\text{Method 1}}(N=36)$ & 1.59 & 3.05 & 4.06 & 2 & 3.55 & 4.52 \\ 
$\Delta_{\hat O}^{\text{Method 2}}(N=36)$ & 1.57 & 3.02 & 4.02 & 1.98 & 3.51 & 4.53   \\ 
 \hline\hline
\end{tabular} 
\end{table}

\subsection{OPE coefficients}
In this subsection, we delve into the finite size correction and error estimation of the OPE coefficients. Generally, the finite size correction arises from the descendants and higher primary operators. For instance, considering $\sigma$, the higher contributions involve $\partial_\mu \sigma$, $(\Box \sigma, \partial_\mu\partial_\nu \sigma)$, $(\partial_\mu\Box \sigma, \partial_\mu\partial_\nu\partial_\rho \sigma)$, $\sigma'$, and so on. The corresponding bulk-defect OPE is given by:
\begin{equation}
G_{\sigma\hat O}(\theta) = \frac{\langle \hat 1|n^z|\hat O\rangle}{\langle \sigma |n^z|1\rangle} \approx \frac{b_{\sigma\hat O}}{(\sin\theta)^{\Delta_\sigma-\Delta_{\hat O}}}\left(1+\frac{c}{R} + \frac{c^\prime}{R^2} +\frac{c^{\prime\prime}}{R^{3}} +O(1/R^{4.180-0.518})\right).
\end{equation}
Similarly, for the higher correction from $\partial_\mu \epsilon$, $T_{\mu\nu}$, $(\Box \epsilon, \partial_\mu\partial_\nu \epsilon)$, $\epsilon'$, and so on, we have:
\begin{equation}
G_{\epsilon\hat O}(\theta) = \frac{\langle \hat 1|n^x|\hat O\rangle-\delta_{\hat O \hat 1}\langle \hat 1|n^x|\hat 1\rangle}{\langle \epsilon |n^x|1\rangle}\approx \frac{b_{\epsilon\hat O}}{(\sin\theta)^{\Delta_\epsilon-\Delta_{\hat O}}}\left(1+\frac{c}{R} + \frac{c^\prime}{R^{3-1.413}} +\frac{c^{\prime\prime}}{R^{2}} +O(1/R^{3.830-1.413})\right).
\end{equation}

Regarding the error bar estimation, we perform the finite size extrapolation using the first two powers to obtain the OPE coefficients $b_{O\hat{O}_n}$. We compare various fitting processes by considering different finite system sizes used in the finite-size extrapolations (the five largest system sizes are always included).   
By comparing the extrapolated OPE coefficients obtained by different fitting processes, we calculate the standard value and  corresponding error bar.

\section{Supplementary Note 3: Attractive fixed point} 
In this section, we aim to demonstrate numerically that the Ising CFT with a line defect
possesses an attractive fixed point under the flow of $h_d$. 

Firstly, we ensure that, under the magnetic line defect the low-energy excitation spectrum is gapless. This can be examined through the scaling analysis, as illustrated in Supplementary Figure \ref{fig:gap}.

\begin{figure}[b]
\includegraphics[width=0.33\linewidth]{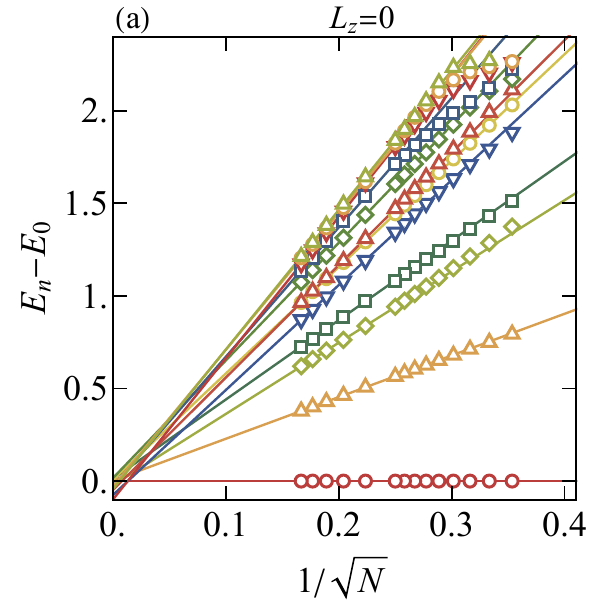}
\includegraphics[width=0.33\linewidth]{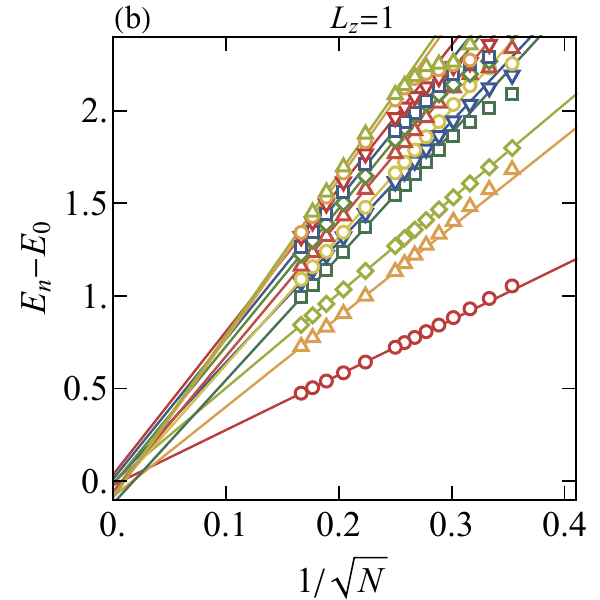}
\caption{\textbf{Energy gap scaling.} The finite-size scaling of energy gap $E_n-E_0$ for the first $10$ states in sector (a)$L_z=0$ and (b)$L_z=1$. System sizes from $N=8$ to $N=15$ (ED) and $N=16$ to $N=36$ (DMRG with $D=5000$). Here we set $h_d=300$.}
\label{fig:gap}
\end{figure}

Second, we examine how the scaling dimensions of primaries converge to the same values for various $h_d$. Finite size corrections arise from irrelevant operators with scaling dimensions $\Delta_{\hat O}>1$. Among these operators, the lowest two primaries are $\hat \phi$ and $\hat \phi'$. Consequently, the finite size correction to the scaling dimension can be approximated as follows up to the first order perturbation:
\begin{equation}\label{seq:correction_firstorder}
\Delta_{\hat O}(N) \approx \Delta_{\hat O}
+\frac{b}{R^{\Delta_{\hat \phi}-1}}
+\frac{c}{R^{\Delta_{\hat \phi'}-1}}
+\text{higher order corrections}.
\end{equation}
By employing this relation and setting $\hat O=\hat \phi,\hat \phi'$, we obtain two consistency equations which give the scaling dimensions $\Delta_{\hat \phi}, \Delta_{\hat \phi'}$. For example, we obtain for $h_d=300$:
\begin{equation}\label{S11}
\Delta_{\hat \phi} \approx 1.63\quad \text{and} \quad \Delta_{\hat \phi'} \approx 3.12.
\end{equation}
One may wonder how the second-order perturbation influences the extrapolated data. We also try to do the finite-size extrapolation using the scaling function by involving the leading second order perturbation
\begin{equation}\label{seq:correction_secondorder}
\Delta_{\hat O}(N) \approx \Delta_{\hat O}
+\frac{b}{R^{\Delta_{\hat \phi}-1}}
+\frac{c}{R^{2(\Delta_{\hat \phi}-1)}}
+\text{higher order corrections}.
\end{equation}
Using Supplementary Supplementary Eq. (\ref{seq:correction_secondorder}), we obtain for $h_d=300$: 
\begin{equation}
\Delta_{\hat \phi} \approx 1.62\quad \text{and} \quad \Delta_{\hat \phi'} \approx 3.08,
\end{equation}
which is almost the same as those obtained in Supplementary Eq. (\ref{S11}). So we conclude that higher-order corrections do not change the extrapolated scaling dimensions qualitatively. 

Subsequently, by applying this approach to various values of $h_d$ ranging from $1$ to $1000$, we observe that the scaling dimensions of $\hat \phi$ and $\hat \phi'$ are insensitive to $h_d$. We examine the displacement operator $\hat D$ in the $L_z=1$ sector. The exact value is $\Delta_{\hat D}=2$, and the numerical results closely match $2$ with very high precision. These findings, the scaling dimensions insensitive to the defect strength $h_d$, provide compelling evidence for the existence of an attractive fixed point in the presence of a line defect. All the results are presented in Supplementary Figure \ref{sfig:deltaBulkVFit}.

\begin{figure}[t]
\includegraphics[width=0.97\linewidth]{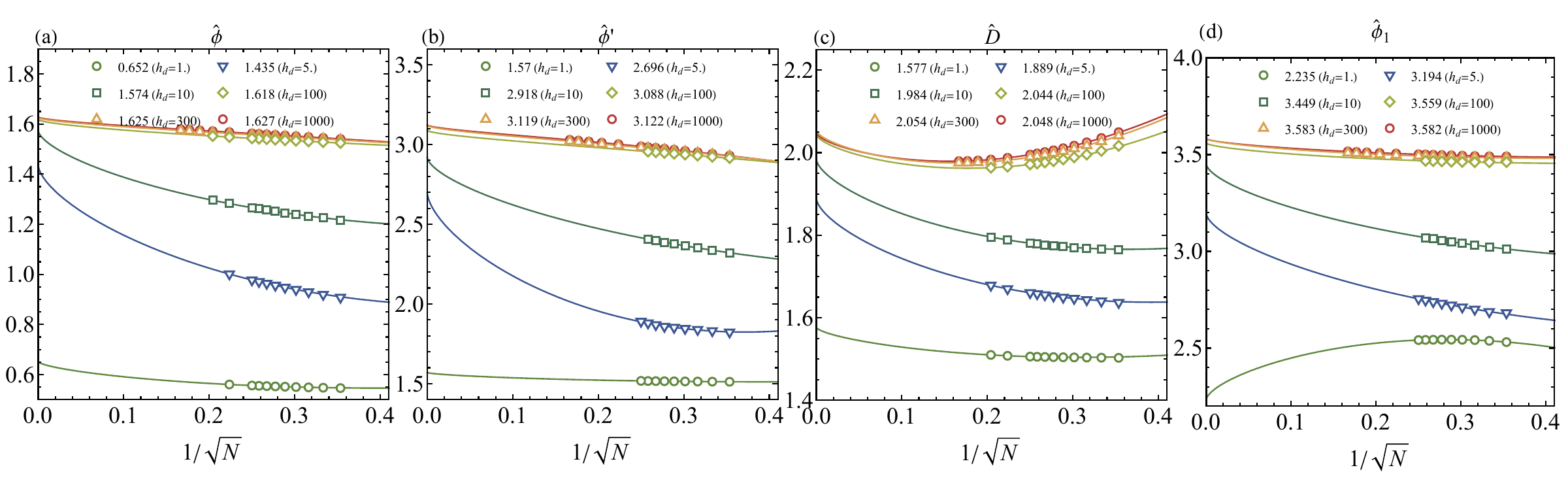}
\caption{\textbf{Finite-size correction of scaling dimensions} for the lower primaries: (a) $\hat \phi$, (b) $\hat\phi'$, (c) $\hat D$, and (d) $\hat \phi_1$, are determined using Supplementary Eq. (\ref{seq:correction_firstorder}). For different values of $h_d$, each primary converges to the same value, indicating the nature of an attractive fixed point. }
\label{sfig:deltaBulkVFit}
\end{figure}

\begin{figure}[b]
\includegraphics[width=0.49\linewidth]{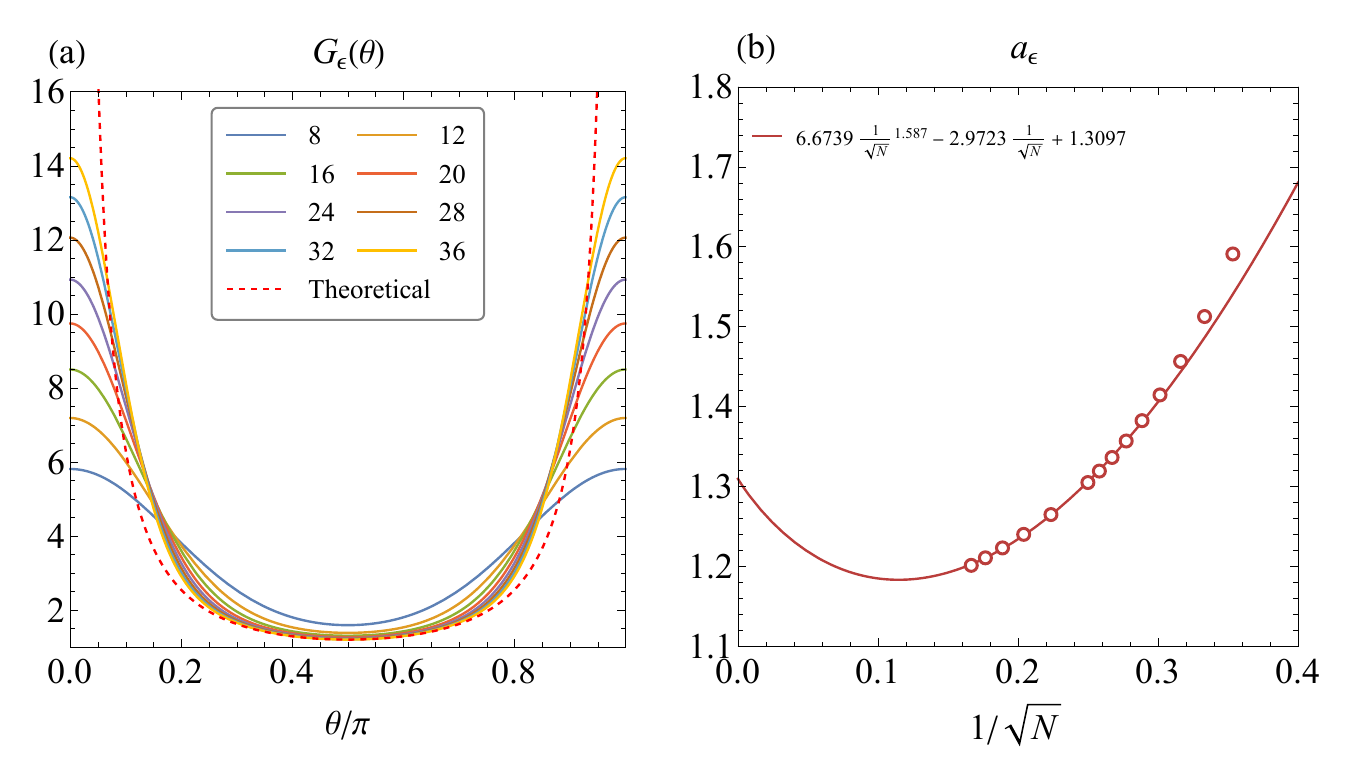}
\includegraphics[width=0.49\linewidth]{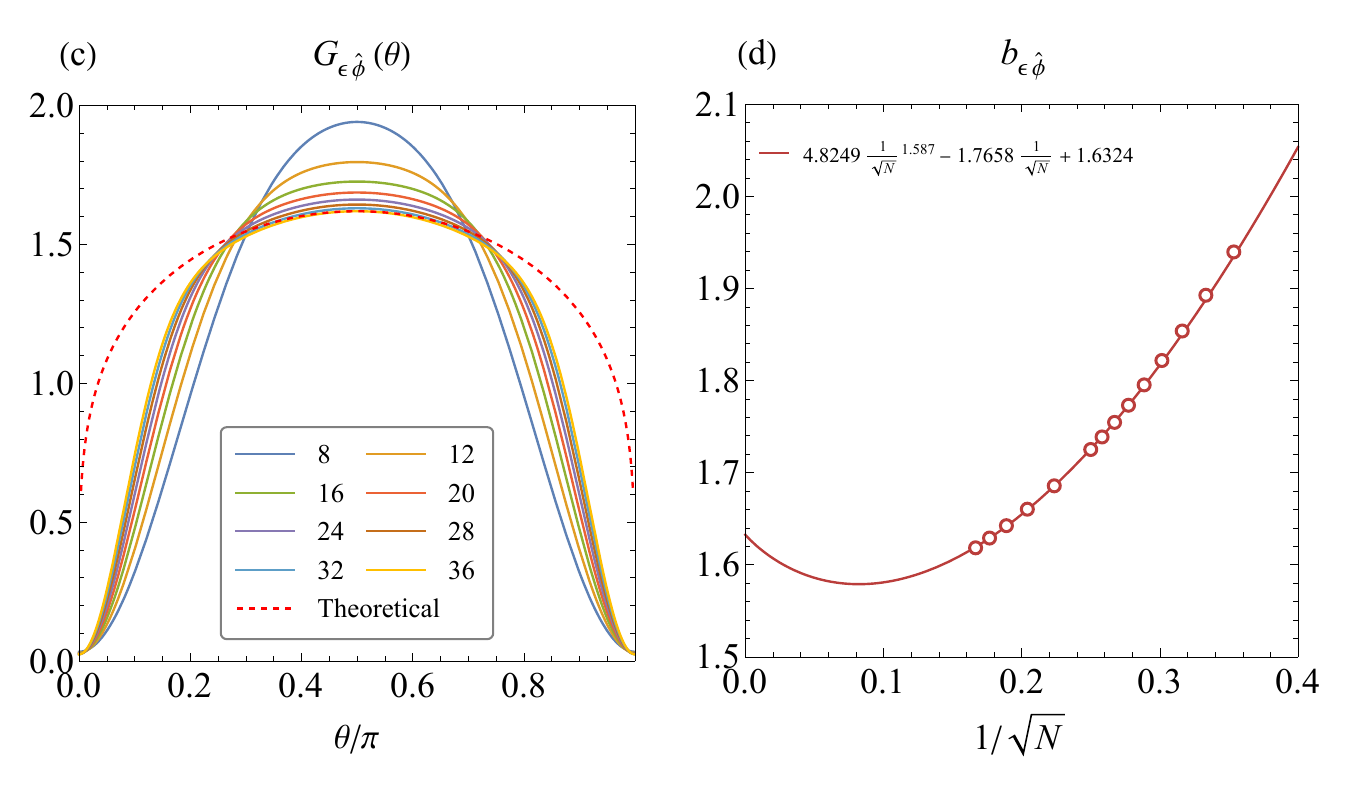}
\caption{\textbf{Correlators related to $\epsilon$} (a-b) The one-point correlator $a_\epsilon$ and its finite-size extrapolation. The result is $a_\epsilon \approx 1.31$. (c-d) The bulk-defect correlator $G_{\epsilon\hat \phi}(\theta)$ with a resulting coefficient of $b_{\epsilon\hat \phi}\approx 1.63$. The dashed lines correspond to the theoretical correlator in Supplementary Eq.~\eqref{seq:OPE} with $b_{\epsilon \hat O}$ and $\Delta_{\hat O}$ from the $N=36$ curves.
The system size ranges from $N=8$ to $N=36$($D=5000$).}
\label{fig:aepsion}
\end{figure}

Finally, we can further consider the limit of $h_d \rightarrow \infty$, as shown in Sec. F. In this limit, we demonstrate that the whole spectra is almost the same with the large $h_d$ regime. 
This indicates that the attractive fixed point is realized in the large $h_d$ limit.

\section{Supplementary Note 4: Numerical results of correlators of $\epsilon$}
In the main text, we have shown the correlators and OPE coefficients related to $\sigma$ (Fig. 4 and related discussion). 
In this section, we present the correlators and OPE coefficients related to the bulk operator $\epsilon$. Following the method described in the main text and in Supplementary References \cite{hu2023operator}, we approximate it using the operator $n^x(\Omega)$. The one-point and bulk-defect OPE are given by:
\begin{equation}
\begin{split}\label{seq:OPE}
G_\epsilon(\theta) &= \frac{\langle \hat 1|n^x|\hat 1\rangle-\langle1|n^x|1\rangle}{\langle \epsilon |n^x|1\rangle} \approx\frac{a_\epsilon}{(\sin\theta)^{\Delta_\epsilon}} + O(1/R)\\
G_{\epsilon\hat O}(\theta) &= \frac{\langle \hat 1|n^x|\hat O\rangle}{\langle \epsilon |n^x|1\rangle}\approx \frac{b_{\epsilon\hat O}}{(\sin\theta)^{\Delta_\epsilon-\Delta_{\hat O}}} + O(1/R)
\end{split}
\end{equation}
Please note that the operator $n^x$ includes an identity component that should be subtracted. Since $\Delta_\epsilon\approx 1.413>0$, $G_\epsilon(\theta)$ should diverge at $\theta=0,\pi$. This behavior is also observed in Supplementary Figure \ref{fig:aepsion}(a). We calculated $G_\epsilon(\theta)$ and extracted the OPE coefficient using $a_\epsilon= G_\epsilon(\pi/2)$. To perform the finite-size extrapolation, we need to consider the finite-size correction given by:
\begin{equation}
G_{\epsilon}(\theta) \approx \frac{1}{(\sin\theta)^{\Delta_\epsilon-\Delta_{\hat O}}}\left( b_{\epsilon\hat O}+\frac{b_{\partial \epsilon\hat O}}{R}+\frac{b_{T_{\mu\nu}\hat O}}{R^{3-1.413}}+O(1/R^2) \right)
\end{equation}
where the first pole $1/R$ comes from $\partial \epsilon$ and the second pole comes from $T_{\mu\nu}$. After the finite-size extrapolation, we find $a_\epsilon\approx 1.31$.

Next, we consider the bulk-defect OPE. Since the lowest primary in dCFT is $\Delta_{\hat \phi}\approx 1.64>\Delta_\epsilon$, the correlator $G_{\epsilon \hat O}(\theta)$ is zero at $\theta=0,\pi$. We present the result for $G_{\epsilon\hat \phi}$ in Supplementary Figure \ref{fig:aepsion}(c), and the finite-size correction follows the same analysis as described above, yielding $b_{\epsilon\hat \phi}\approx 1.63$.

\section{Supplementary Note 5: Ward identity regarding to the displacement operator}
In dCFT, the displacement operator is related to the stress tensor in the bulk CFT, $ \partial_\mu T_{\mu\nu}(x_\perp, x_\parallel) =  \delta(x_\perp) \hat D_\nu(x_\parallel) $. The stress-tensor has a canonical normalization through the Ward-identity, so the normalization of $\hat D$ is also fixed. Therefore, the two-point correlator of displacement operator will not be normalized to $1$, instead, it is 
\begin{equation}
\langle \hat D_i(x)\hat D_j(0)\rangle = C_{\hat D}\frac{\delta_{ij}}{x^4},
\end{equation}
where the normalization factor $C_{\hat D}$ is the Zamolodchikov norm \cite{Billo2016defect} or central charge. For this canonically normalized displacement operator, its bulk-defect OPE coefficients are constrained by the Ward identity \cite{Billo2016defect}:
\begin{equation}
\Delta_O a_O = \frac{\pi}{2} b_{O\hat D},
\end{equation}
where $O$ is a bulk scalar primary operator. $ b_{O\hat D}$ is a bulk-defect OPE coefficient defined by the two-point correlator in $\mathbb R^3$, 
\begin{equation}
\langle O(x) \hat D_i(0) \rangle =b_{O \hat D} \frac{ x_{\perp,i}}{|x_\perp|^{\Delta-1} |x|^4 }.
\end{equation}

In our fuzzy sphere computation, we can directly use the state of displacement operator to compute the correlators involving the displacement operator. Specifically, we have state with well defined $SO(2)\cong U(1)$ quantum number $L_z$, 
\begin{equation}
|\hat D, L_z=\pm 1\rangle = \frac{1}{\sqrt{2C_{\hat D}}} (\hat D_1(\tau = -\infty) \pm i \hat D_2(\tau = -\infty)) | \hat 1 \rangle.
\end{equation}
The numerical factor $1/\sqrt{2C_{\hat D}}$ is chosen such that $\langle \hat D, L_z=1 | \hat D, L_z=1 \rangle=\langle \hat D, L_z=-1 | \hat D, L_z=-1 \rangle=1$. On the cylinder $S^2\times \mathbb R$, we have
\begin{equation}
\langle \hat{1} | O(\tau=0, \theta,\varphi) |\hat D, L_z=1\rangle = \frac{b_{O\hat D}}{\sqrt{2 C_{\hat D}}}  \frac{e^{i\varphi} }{R^\Delta (\sin\theta)^{\Delta-1}} 
\end{equation}
Therefore, we can extract $C_{\hat D}$ using
\begin{equation}\label{seq:CC}
\sqrt{2C_{\hat D}} = \frac{2}{\pi} \frac{\Delta_O G_{O}(\theta=\pi/2)}{ G_{O\hat D}(\theta=\pi/2) },
\end{equation}
where,
\begin{equation}
G_{O\hat{D}}\equiv \frac{  e^{-i\varphi} \langle \hat{1} | O(\tau=0, \theta,\varphi) |\hat D, L_z=1\rangle}{\langle 1| O(\tau=0, \theta,\varphi) |O\rangle} ,
    \end{equation}
 The results are depicted in Supplementary Figure \ref{fig:bsigmaD}(c). For comparison, we calculate the results by setting $O=\sigma$ and $O=\epsilon$, yielding $C_{\hat D}=0.27(1)$ for $O=\sigma$ and $C_{\hat D}=0.30(8)$ for $O=\epsilon$.

\begin{figure}[tb]
\includegraphics[width=0.25\linewidth]{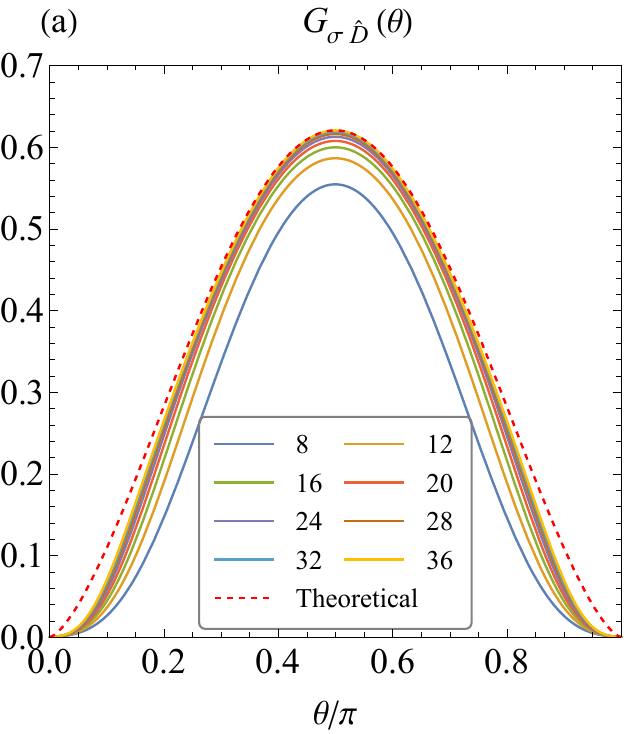}
\includegraphics[width=0.25\linewidth]{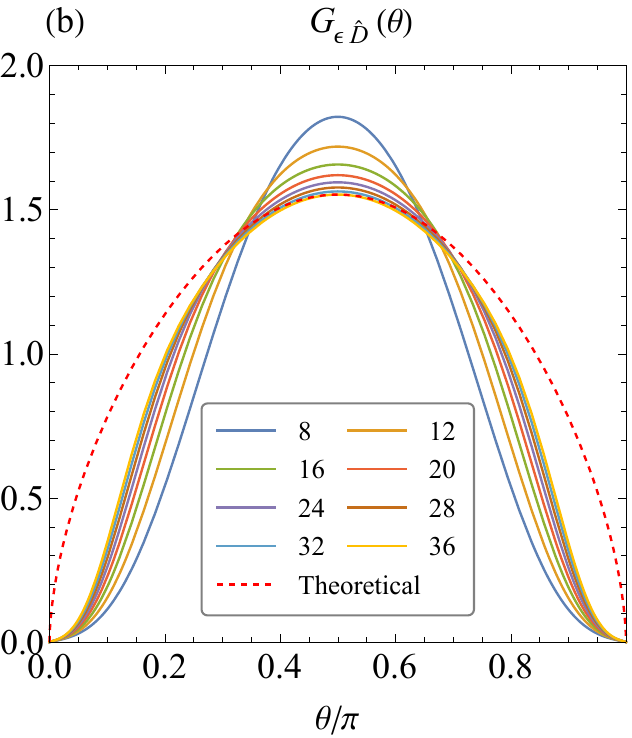}
\includegraphics[width=0.26\linewidth]{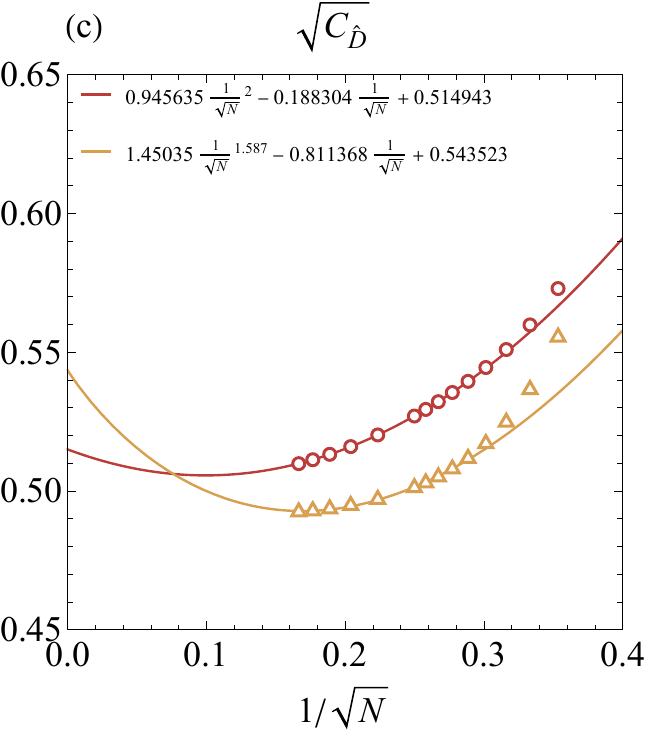}
\caption{Correlators between the displacement operator and bulk operator: (a) $O=\sigma$ and (b) $O=\epsilon$. The dashed lines correspond to the theoretical correlator in Supplementary Eq. (\ref{seq:OPE}) with $b_{\epsilon \hat D}$ and $\Delta_{\hat D}$ from the $N=36$ curves. (c) The extracted Zamolodchikov norm from Supplementary Eq. (\ref{seq:CC}). The red circles represent the results obtained when $O=\sigma$, while the yellow triangles correspond to the results when $O=\epsilon$.} 
\label{fig:bsigmaD}
\end{figure}

\section{Supplementary Note 6: In the limit of $h_d\rightarrow  \infty$}
In this section, we will consider whether the system remains at the dCFT fixed point when the defect strength $h_d$ tends to infinity. 
First, we consider the defect term
\begin{equation}
\begin{split}
    H_d &= 2\pi h_d \left[ n^z(\theta=0,\varphi=0)+n^z(\theta=\pi,\varphi=0)  \right]\\
    &= 2\pi\sum_l \left[ Y_{l,0}(\theta=0,\varphi=0)+Y_{l,0}(\theta=\pi,\varphi=0) \right]n^z_{l,0}\\
    &=\sum_{m_1} \left[C_{m_1}(\theta=0)+C_{m_1}(\theta=\pi)\right]
    \hat c_{m_1}^\dagger\sigma^z\hat c_{m_1}
\end{split}
\end{equation}
where $C_{m_1}(\theta)$ is 
\begin{equation}
    C_{m_1}(\theta) = \frac12 h_d \pi \sum_l (-1)^{3s+m_1+l} P_l(\cos\theta) (2l+1) \tj{s}{s}{l}{s}{-s}{0} \tj{s}{s}{l}{m_1}{-m_1}{0} 
\end{equation}
The Legendre polynomial $P_l(x)$ have special values $P_l(1) = 1$ and $P_l(-1)=(-1)^l$, substituting into the last equation and using the orthogonal relation of $3j-$symbol, we have
\begin{equation}
    C_{m_1}(\theta=0) = \frac12 h_d \delta_{m_1,-s}\qquad
    C_{m_1}(\theta=\pi) = \frac12 h_d \delta_{m_1,s}.
\end{equation}
Thus, the defect term has a simplified form in orbital space (see also \cite{zhou2024gfunction})
\begin{equation}
    H_d = \frac 12 h_d \left( \hat c_{s}^\dagger\sigma^z\hat c_{s}+\hat c_{-s}^\dagger\sigma^z\hat c_{-s} \right).
\end{equation}
In orbital space, the defect term only acts on the $m=\pm s$ orbitals. This implies that if the defect strength $h_d$ is sufficiently large, we can fix the spins of the $m=\pm s$ orbitals to $\downarrow$ in numerical computations and then optimize the other orbitals. This approach brings two advantages: firstly, it reduces the number of orbitals, which can enhance computational efficiency; secondly, it avoids numerical difficulties arising from excessively large $h_d$.

Using the method described in Supple. Mat. Sec. B, we similarly computed the spectra for $N$ ranging from 8 to 36 at $h_d=\infty$ and plotted them in Supplementary Figure \ref{smfig:inftower}. We can observe that the system still exhibits conformal symmetry, indicating it remains at the fixed point of dCFT. Additionally, we list the scaling dimensions of several lower primary fields in Supplementary Table \ref{smtab:infprimary}. A comparison with the finite $h_d=300$ case mentioned in the main text reveals consistency within the error range. This indicates that the choice of $h_d=300$ in the main text is sufficiently large. Combining this observation with the discussion in the main text, we can confidently assert the existence of an attractive dCFT fixed point at $h_d=\infty$ for the 3D Ising critical point under the imposition of the magnetic field line.

\begin{figure}[t]
\includegraphics[width=0.33\linewidth]{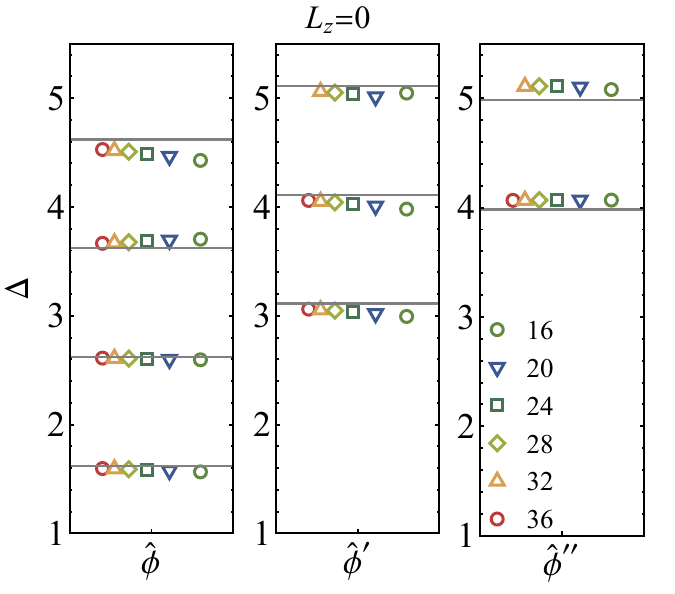}
\includegraphics[width=0.33\linewidth]{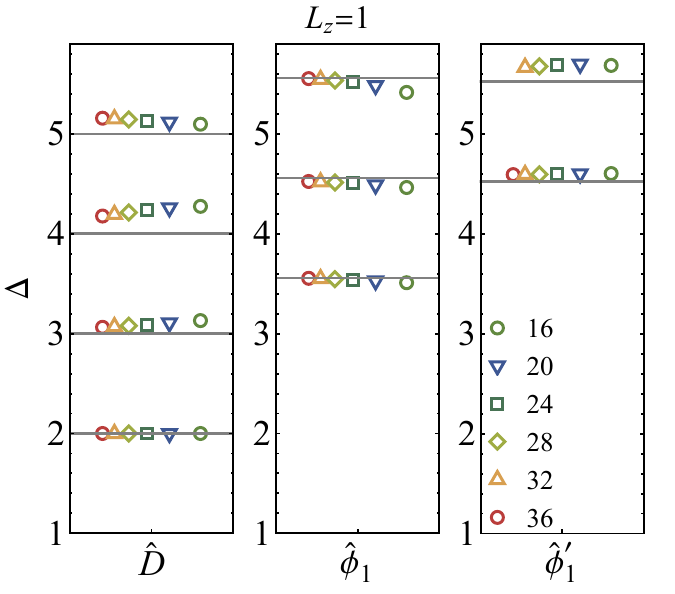}
\caption{Conformal tower of defect primaries in $h_d\rightarrow\infty$. Defect primary fields and their descendants with global symmetry (a) $L_z=0$ and (b) $L_z=1$. 
The grey horizon lines stand for the theoretical expectation for descendants.
Different colored symbols represent the results based on various system sizes. By increasing system size $N$ all of the scaling dimensions approach the theoretical values consistently, supporting an emergent conformal symmetry in the thermodynamic limit.}
\label{smfig:inftower}
\end{figure}

\begin{table}
\setlength{\tabcolsep}{0.2cm}
\renewcommand{\arraystretch}{1.4}
    \centering
    \caption{The scaling dimensions of primary fields under infinite and finite defect strengths, are determined through the state-operator correspondence on the fuzzy sphere. 
    Please see a detailed analysis of errors and finite-size extrapolation in Supple. Mat. Sec. B-C.}\label{smtab:infprimary}
\begin{tabular}{c|ccc|ccc} 
\hline\hline
&\multicolumn{3}{c|}{$L_z=0$}& \multicolumn{3}{c}{$L_z=1$} \\
    \hline
&$\hat \phi$ & $\hat \phi'$ & $\hat \phi''$  & $\hat D$ & $\hat \phi_1$ & $\hat \phi_1'$\\    
 $h_d=\infty$&1.63(4) & 3.12(9) & 4.04(3) & 2.05(7) & 3.58(7) &  4.62(8)  \\  
 $h_d=300$&1.63(6) & 3.12(10) & 4.06(18) & 2.05(7) & 3.58(7) &  4.64(14)  \\   
 \hline\hline
\end{tabular} 
\end{table}


\end{document}